\documentclass[notitlepage,twocolumn,superscriptaddress,longbibliography,aps,nofootinbib]{revtex4-1}

\def\@bibdataout@aps{%
\immediate\write\@bibdataout{%
@CONTROL{%
apsrev41Control%
\longbibliography@sw{%
    ,author="08",editor="1",pages="1",title="0",year="1"%
    }{%
    ,author="08",editor="1",pages="1",title="",year="1"%
    }%
  }%
}%
\if@filesw \immediate \write \@auxout {\string \citation {apsrev41Control}}\fi 
}
\makeatother

\usepackage{graphicx}
\usepackage{tikz}

\usepackage{xcolor}

\usepackage{dcolumn}
\usepackage{bm}
\usepackage[breaklinks=true,colorlinks=true,linkcolor=blue,urlcolor=blue,citecolor=blue,pagebackref=false]{hyperref}

\usepackage{epsf,epsfig,amsmath,amssymb,dsfont}
\usepackage{amsthm,mathrsfs} 
\usepackage{multirow}
\usepackage{float}
\usepackage{array}
\usepackage{mathtools}
\usepackage[normalem]{ulem}

\usepackage[caption=false]{subfig}

\allowdisplaybreaks

\newcommand{\integral}[3]{\int_{#2}^{#3} \!\! \mathrm{d} #1 \,}
\newcommand{\difffrac}[2]{\frac{\mathrm{d} #1}{\mathrm{d} #2}}
\newcommand{\diffffrac}[3]{\frac{\mathrm{d}^{#3} #1}{{\mathrm{d} #2}^{#3}}}

\newcommand{\diff}{\text{d}}

\newcommand{\sign}{\text{sign}}

\newcommand{\ii}{\mathrm{i}}
\newcommand{\ee}[1]{\mathrm{e}^{#1}} %
\renewcommand{\Im}{\ensuremath{\mathrm{Im}}}
\renewcommand{\Re}{\ensuremath{\mathrm{Re}}}

\newcommand{\exptvalstate}[2]{\left<\vphantom{#1}#2 \right|#1\left|\vphantom{#1}#2 \right>}

\newcommand{\Wmanpsix}[3]{W^{#3}\left(#1;#2\right) }

\newcommand{\ket}[1]{\left| {#1} \right\rangle}
\newcommand{\bra}[1]{\left\langle {#1} \right|}

\newcommand{\ketbra}[2]{\left| {#1}\vphantom{#2} \right\rangle\!\left\langle {#2}\vphantom{#1} \right|}

\newcommand{\da}{\textsc{a}}
\newcommand{\db}{\textsc{b}}
\newcommand{\dd}{\textsc{d}}

\newcommand{\Hit}[2]{H^{#2}_{\mathrm{int},\,#1}}

\newcommand{\coord}[1]{\mathsf{#1}}
\newcommand{\spacoord}[1]{\bm{#1}}

\newcommand{\indmode}{\ell\omega} %
\newcommand{\indmodem}{\ell,-\omega} %
\newcommand{\Glw}[1]{G_{\indmode}^{#1}}

\newcommand{\Rin}{R^{\text{in}}_{\indmode}}
\newcommand{\Rup}{R^{\text{up}}_{\indmode}}
\newcommand{\RinC}{\bar{R}^{\text{in}}_{\indmode}}
\newcommand{\RupC}{\bar{R}^{\text{up}}_{\indmode}}

\newcommand{\RinmC}{\bar{R}^{\text{in}}_{\indmodem}}
\newcommand{\RupmC}{\bar{R}^{\text{up}}_{\indmodem}}
\newcommand{\Rinup}{R^{\text{in/up}}_{\indmode}}
\newcommand{\RinupC}{\bar{R}^{\text{in/up}}_{\indmode}}

\newcommand{\RinccC}{\bar{R}^{\text{in}^*}_{\indmode}}
\newcommand{\RupccC}{\bar{R}^{\text{up}^*}_{\indmode}}

\newcommand{\nn}{\nonumber\\}

\newcommand{\erfi}{\mathrm{erfi}}
\newcommand{\erfc}{\mathrm{erfc}}

\newcommand{\ie}{{\it i.e.},\ }

\AtBeginDocument{%
    \newwrite\bibnotes
    \def\bibnotesext{Notes.bib}
    \immediate\openout\bibnotes=\jobname\bibnotesext
    \immediate\write\bibnotes{@CONTROL{REVTEX41Control}}
    \immediate\write\bibnotes{@CONTROL{%
    apsrev41Control,author="08",editor="1",pages="1",title="0",year="1"}}
     \if@filesw
     \immediate\write\@auxout{\string\citation{apsrev41Control}}%
    \fi
}%

\begin{document}

\title{Lensing of Vacuum Entanglement near Schwarzschild Black Holes\\
  } 
\author{Jo\~ao G. A. Carib\'{e}}
\affiliation{Centro Brasileiro de Pesquisas F\'isicas (CBPF), Rio de Janeiro, 
CEP 22290-180, 
Brazil.}
\author{Robert H. Jonsson}
\affiliation{Max Planck Institute of Quantum Optics, Hans-Kopfermann-Str.~1, 85748 Garching, Germany}
\affiliation{Nordita, KTH Royal Institute of Technology and Stockholm University, Hannes Alfv\'ens v\"ag~12, SE-106 91 Stockholm, Sweden}
\author{Marc Casals}
\affiliation{Institut f\"ur Theoretische Physik, Universit\"at Leipzig,  Br\"uderstra\ss e ~16, 04103 Leipzig, Germany}
\affiliation{Centro Brasileiro de Pesquisas F\'isicas (CBPF), Rio de Janeiro, 
CEP 22290-180, 
Brazil.}
\affiliation{School of Mathematics and Statistics, University College Dublin, Belfield, Dublin 4, Ireland.}
\affiliation{Laboratoire Univers et Th\'eories, Observatoire de Paris, CNRS, Universit\'e PSL,
Universit\'e de Paris, 92190 Meudon, France}
\author{Achim Kempf}
\affiliation{Institute for Quantum Computing, University of Waterloo, Waterloo, Ontario, N2L 3G1, Canada}
\affiliation{Department of Physics \& Astronomy, University of Waterloo, Waterloo, Ontario, N2L 3G1, Canada}
\affiliation{Perimeter Institute for Theoretical Physics, 31 Caroline St N, Waterloo, Ontario, N2L 2Y5, Canada}
\affiliation{Department of Applied Mathematics, University of Waterloo, Waterloo, Ontario, N2L 3G1, Canada}
\author{Eduardo Mart\'{i}n-Mart\'{i}nez}
\affiliation{Institute for Quantum Computing, University of Waterloo, Waterloo, Ontario, N2L 3G1, Canada}
\affiliation{Department of Physics \& Astronomy, University of Waterloo, Waterloo, Ontario, N2L 3G1, Canada}
\affiliation{Perimeter Institute for Theoretical Physics, 31 Caroline St N, Waterloo, Ontario, N2L 2Y5, Canada}
\affiliation{Department of Applied Mathematics, University of Waterloo, Waterloo, Ontario, N2L 3G1, Canada}

\begin{abstract}

An important feature of Schwarzschild spacetime is the presence of  
orbiting null geodesics
and caustics. Their presence implies strong gravitational lensing effects for matter and radiation, i.e., for excitations of quantum fields. Here, we raise the question whether the lensing manifests itself also in the vacuum of quantum fields, namely by lensing the distribution of vacuum entanglement.
To explore this possibility, we use the method of entanglement harvesting, where initially unentangled localized quantum systems  are temporarily coupled to the field at different locations. 
We find that for the Boulware, Hartle-Hawking and Unruh vacua in 3+1-dimensional Schwarzschild spacetime, the harvesting of  vacuum entanglement is indeed greatly amplified  near caustics. In particular, we establish that preexisting vacuum entanglement can also be harvested for lightlike separations.
\end{abstract}

\maketitle

\section{Introduction}
The presence of entanglement between spatially separated degrees of freedom of a quantum field is a basic phenomenon that occurs even for free fields in the vacuum state on a flat background spacetime \cite{vacuumEntanglement,vacuumBell}. The origin of this vacuum entanglement can be traced back to the fact that, in wave equations, neighboring field oscillators must be coupled to each other in order to describe the propagation of waves.
The coupling between the neighboring field oscillators is through spatial derivatives in the wave operators, such as the d'Alembertian and Dirac operator. %
It is this coupling between neighboring field oscillators that also causes the ground state of the local field oscillators to be an entangled state. 

In curved spacetimes,  curvature impacts the derivatives in the wave operators which then impact the  entanglement in the field. Therefore, curvature also impacts the field correlations. Conversely, it has been shown that the imprint that curvature leaves in the field correlators is actually complete in the sense that the metric can be reconstructed from the field correlators \cite{saravani2016spacetime,kempf2021replacing}. As was shown in Ref.~\cite{TalesEduSpacetimefromCorrelators}, the metric can also be reconstructed from the correlations between local measurements of the field. 
The entanglement structure of  quantum fields plays a fundamental role in investigations of phenomena from holography to Hawking radiation and the black hole information loss problem~\cite{BenensteinEntropy1973,SorkinArea1986,areaLaw1993,areaLawReview2010,witten,HawkingRadiation,hawking_particle_1975,Davies1974,Page1993,Page2013,Penington2020}.

To probe the spacetime distribution of entanglement in a quantum field, a versatile method is to couple initially unentangled localized quantum systems to the field at different spacetime regions. The amount of entanglement that the localized systems acquired can be determined by standard methods~\cite{reznik1,Pozas-Kerstjens:2015}. 

When the entanglement acquired from the field by localized quantum systems is extracted from the entanglement that was preexisting in the field, this protocol has become known as entanglement harvesting. %
Entanglement harvesting has been investigated in a number of scenarios since first hinted at in Refs.~\cite{Valentini1991,Reznik2003},  in both flat and curved spacetime. It has been proven that  entanglement harvesting can capture the geometry~\cite{Nick} and topology~\cite{topology} of the underlying spacetime. So far, the scenario where entanglement is harvested from the field in the presence of black holes has only been studied in very idealized scenarios such as 2+1-dimensional (Ba\~nados-Teitelboim-Zanelli) black holes~\cite{Henderson_2018} and 1+1-dimensional  spacetimes with horizons~\cite{KenMannSCHBH}. The question of entanglement harvesting near a  black hole in 3+1 spacetime dimensions has remained 
open. However, this case is of particular interest since one can expect new phenomenology, for example, due to lensing (see Ref.~\cite{cliche2011vacuum}) and due to the fact that orbiting null geodesics and caustics can connect the two localized harvesting systems.

The phenomenology arising from orbiting null geodesics and caustics on communication through quantum fields close to a Schwarzschild black hole was addressed in Ref.~\cite{jonssonCommunicationQuantumFields2020}.
Here, we investigate in detail entanglement harvesting in a four-dimensional Schwarzschild spacetime, for the cases of the Boulware, Hartle-Hawking and Unruh vacua, using tools similar to those applied   in Ref.~\cite{jonssonCommunicationQuantumFields2020}. We take the localized quantum systems to be static localized two-level quantum systems with a monopole coupling to a Klein-Gordon field, i.e., so-called Unruh-DeWitt (UDW) detectors, or detectors for short.
Within this setup, we analyze the impact of the presence of caustics and of the fact that the detectors can be connected by secondary null geodesics, including the case where detectors are placed close to the event horizon. 

We find, that the presence of caustics alters the characteristics of entanglement harvesting in two  particular ways in comparison to flat spacetime. First, due to a lensing-like effect caused by the focusing of null geodesics, the final entanglement between detectors can become greatly amplified near the caustics in comparison to comparably placed detectors away from caustics. Second, due to changes in the singularity structure of the Wightman function which happen when the field waves cross through caustics, we observe that timelike-separated detectors can harvest preexisting entanglement from the field if they are aligned along secondary null geodesics, \ie null geodesics which orbit half of the black hole and so they have passed through one caustic.
To the best of our knowledge, such harvesting of preexisting entanglement for lightlike separations has never been observed before in any spacetime.

Sec.~\ref{sec:Wman_sing_structure} begins by discussing the treatment and properties of the Wightman function of a massless scalar field on Schwarzschild spacetime and, in particular, its global singularity structure.
Sec.~\ref{sec:harvesting} introduces the detector model, its perturbative treatment and negativity as entanglement measure for the detectors' final state. Sec.~\ref{sec:whenisharvestharvest} discusses when the entanglement between detectors can be attributed to the harvesting of preexisting entanglement from the field. Sec.~\ref{sec:results_harvesting} presents our actual results for specific detectors and the article closes in Sec.~\ref{sec:outlook} with a discussion and outlook.
The appendices collect supplemental figures. Furthermore, App.~\ref{app:LM_expressions} gives the calculations of the perturbative contributions to the detectors' state and App.~\ref{app:num techniques} discusses the numerical techniques for the evaluation of the Wightman functions and integrals evolving the Wightman function.

We use the natural units in which $c=G=\hbar=1$.

\section{Wightman function in Schwarzschild spacetime}

In this section we introduce a quantum scalar field in Schwarzschild spacetime as well as the Wightman function when the scalar field is in three quantum states of interest (namely, Boulware, Unruh and Hartle-Hawking). 
The reader familiar with these may wish to skip to Sec.~\ref{sec:Wman_sing_structure}, reviewing literature results on the global singularity structure of the Wightman function as the field wave front passes through caustics of Schwarzschild spacetime which motivate our search for gravitational lensing of entanglement harvesting.

\subsection{Klein-Gordon quantum field in Schwarzschild spacetime}
In this section, we briefly review the treatment of a massless Klein-Gordon field in Schwarzschild spacetime, 
and the expressions for the Wightman function of the three field states we consider, which are needed for the perturbative treatment of the field-detector interaction.

The line element of the outer region of Schwarzschild spacetime in Schwarzschild coordinates
$\{t\in\mathbb{R},\, r\in (2M,\infty),\, \theta\in [0,\pi],\, \varphi\in [0,2\pi)\}$, is given by
\begin{equation}
\diff s^2=-f(r) \diff t^2+f(r)^{-1}\diff r^2+r^2\left(\diff\theta^2+\sin^2\theta \diff\varphi^2\right),\label{eq:schwarzschild_metric}
\end{equation}
where $f(r):=1-2M/r$ $M$ is the mass of the black hole and $r=2M$ is the radius of the event horizon. 
In this outer region, we consider a scalar quantum field $\hat\phi$ obeying the Klein-Gordon (K-G) equation $\Box\hat\phi = 0$.
Considering the Schwarzschild spacetime and availing of its spherical symmetry, a general real-valued solution for that equation can be written as 
\begin{widetext}
    \begin{equation}\label{eq:field exp}
        \hat\phi\left(\coord{x}\right) = \sum_{\ell = 0}^{\infty}\sum_{m = -\ell}^{\ell}\integral{\omega}{0}{\infty}\left(\hat\alpha^{\mathrm{in}}_{\ell m \omega}\phi^{\mathrm{in}}_{\ell m \omega}\left(\coord{x}\right)+\hat\alpha^{\mathrm{up}}_{\ell m \omega}\phi^{\mathrm{up}}_{\ell m \omega}\left(\coord{x}\right) + \hat\alpha^{\mathrm{in}\dagger}_{\ell m \omega}\phi^{\mathrm{in}*}_{\ell m \omega}\left(\coord{x}\right)+\hat\alpha^{\mathrm{up}\dagger}_{\ell m \omega}\phi^{\mathrm{up}*}_{\ell m \omega}\left(\coord{x}\right)\right),
    \end{equation}
\end{widetext}
where $\coord{x}$ denotes a spacetime point, $\hat\alpha^{\mathrm{in/up}\dagger}_{\ell m \omega}$ are creation and annihilation operators and
\begin{equation}\label{eq:in_up_field_modes}
    \phi^{\mathrm{in/up}}_{\ell m \omega}\left(t,r,\theta,\varphi\right) = \frac{1}{\sqrt{4\pi\omega}}e^{-\ii\omega t}Y_{\ell m}\left(\theta,\varphi\right)\frac{\Rinup\left(r\right)}{r},
\end{equation}
with $Y_{\ell m}$ being the spherical harmonic of degree $\ell$ and order $m$ and $\Rinup$ are radial factors. 

Substituting the field modes \eqref{eq:in_up_field_modes} into the K-G equation leads to the conclusion that $\Rinup$ obeys
\begin{equation}
   \diffffrac{\Rinup}{r_*}{2}+\left(\omega^2-V_{\ell}\left(r\right)\right)\Rinup=0,\label{eq:regge_wheeler}
\end{equation}
where $r_* = r + 4M\ln\left\vert\frac{r}{2M}-1\right\vert$$\in (-\infty,+\infty)$ is the tortoise coordinate and $V_{\ell}\left(r\right)=f(r) \left(\frac{2M}{r^3}+\frac{\ell\left(\ell+1\right)}{r^2}\right)$  is an effective potential.

The two linearly independent solutions $\Rin$ and $\Rup$ are defined by the asymptotic boundary conditions 
\begin{align}
    &\Rin \sim \begin{cases}
        \ee{-\ii\omega r_*},&r_*\to -\infty,\\
        I_{\ell \omega}\ee{-\ii\omega r_*} + \rho^{\mathrm{in}}_{\ell \omega}\ee{\ii\omega r_*},&r_*\to \infty,\\
    \end{cases}\\
    &\Rup \sim \begin{cases}
        I_{\ell \omega}\ee{\ii\omega r_*} + \rho^{\mathrm{up}}_{\ell \omega}\ee{-\ii\omega r_*},&r_*\to -\infty,\\
        \ee{\ii\omega r_*},&r_*\to \infty.
    \end{cases}
\end{align}
Here, $\rho^{\mathrm{in/up}}_{\ell \omega}\in\mathbb{C}$ are the reflection amplitudes and $I_{\ell \omega}\in\mathbb{C}$ is the incidence amplitude. For  compatibility with Ref.~\cite{candelasVacuumPolarizationSchwarzschild1980} and ease of notation, let us also define
\begin{equation}\label{eq:defn_irnupc}
    \RinupC \equiv \frac{1}{r I_{\ell \omega}}\Rinup.
\end{equation}
Note that it suffices to calculate the modes for just $\omega\ge 0$ 
and use the following symmetries for $\omega<0$:
\begin{equation}\label{eq:rad symms}
\RinccC=\RinmC, \quad \RupccC=\RupmC.
\end{equation}

Calculating the response of the detectors to the interaction with the quantized field $\hat\phi$ requires, as we shall see in the next section, the Wightman function.
The Wightman function, when the quantum field %
is in a state $\ket{\Psi}$, is defined as
\begin{equation}
\Wmanpsix{\coord{x}}{\coord{x}'}\Psi\equiv \bra\Psi\hat\phi(\coord{x})\hat\phi(\coord{x}')\ket{\Psi}.
\end{equation}
It is thus a two-point function satisfying the homogeneous K-G equation.
Henceforth we shall only consider quantum states $\ket{\Psi}$ in regions of spacetime where they satisfy the Hadamard property (namely, Eq.~\eqref{eq:Wightman Had}, is satisfied for the Wightmant function in such states).
The Wightman function in Schwarzschild spacetime is given by~\cite{candelasVacuumPolarizationSchwarzschild1980}
\begin{equation}\label{eq:GF}
\begin{split}
&\Wmanpsix{\coord{x}}{\coord{x}'}\Psi=
\\&\quad
\frac{1}{\left(4\pi\right)^2}\sum_{\ell=0}^{\infty}(2\ell+1)P_{\ell}(\cos\gamma)\int_{-\infty}^{\infty}\frac{d\omega}{\omega}
e^{-i\omega \Delta t}
\Glw{\Psi}(r,r';\Delta t),
\end{split}
\end{equation}
where $\gamma$ is the angular separation between the  two spacetime points $\coord x$ and $\coord{x'}$
and $\Delta t\equiv t-t'$. %
The integral kernel $\Glw{\Psi}$ depends on the quantum state of the field. For the Boulware~\cite{boulware1975spin} ($\Psi=B$), Unruh~\cite{Unruh1976} ($\Psi=U$) and Hartle-Hawking~\cite{hartle1976path} states ($\Psi=H$) it takes the forms
\begin{align}\label{eq:FGF modes B}
\Glw{B}&=
\theta(\omega)\left(\RupC(r)\RupccC(r')+\RinC(r)\RinccC(r') \right),
\\
\label{eq:FGF modes U}\Glw{U}&=
\frac{
\RupC(r)\RupccC(r')
}
{1-e^{-2\pi\omega/\kappa}}
+\theta(\omega)\RinC(r)\RinccC(r'), 
\\
\label{eq:FGF modes H}
\Glw{H}&=
\frac{\RupC(r)\RupccC(r')
+\RinccC(r)\RinC(r')}{1-e^{-2\pi\omega/\kappa}},
\end{align}
where $\kappa\equiv 1/(4M)$ is the surface gravity, and $\RinupC$ denotes the (rescaled) radial factor of the ingoing and upgoing solutions to the wave equation, as defined in Eq.~\eqref{eq:defn_irnupc}.
App.~\ref{app:num techniques}  discusses the numerical techniques used for the evaluation of the Wightman function.

The expressions above convey that in the Boulware state both ingoing and upgoing modes are in their ground state, whereas in the Unruh state the upgoing modes are thermalized, and in the Hawking state both ingoing and upgoing modes are thermalized.

\subsection{Singularity structure of the Wightman function}
\label{sec:Wman_sing_structure}

The Hadamard form for the Wightman  function $\Wmanpsix{\coord{x}}{\coord{x}'}\Psi$ is an analytic expression which is defined in a local neighborhood\footnote{More precisely, the Hadamard form is only defined 
in a normal neighborhood of $\coord{x}$: 
a neighbourhood $\mathcal{N}(\coord x)$ of $\coord x$ such that every $\coord{x}’ \in \mathcal{N}(\coord x)$ is connected to $\coord x$ by a {\it unique} geodesic which lies in $\mathcal{N}(\coord x)$.} of the source  point $\coord{x}$
and explicitly shows its singularity structure. Explicitly, it is (e.g., Refs.~\cite{HOLLANDS20151,DeWitt:1960,Hadamard})
\begin{equation}\label{eq:Wightman Had}
\begin{split}
&
\Wmanpsix{\coord{x}}{\coord{x}'}\Psi=
\\& \quad
\lim_{\epsilon \rightarrow 0^+}
\frac{1}{4\pi^2}
\left[\frac{u}{\sigma+\ii\, \epsilon\, \Delta t}-v\ln\left(\sigma+\ii\,\epsilon\, \Delta t\right)+w\right],
\end{split}
\end{equation}
where  $u=u(\coord{x},\coord{x}')$, $v=v(\coord{x},\coord{x}')$ and $w=w(\coord{x},\coord{x}')$
are regular and real-valued biscalars. The so-called Synge's world function $\sigma=\sigma(\coord{x},\coord{x}')$ is 
 equal to one-half of the square of the  geodesic distance joining $\coord{x}$ and $\coord{x}'$,
 which implies that $\sigma$
 is negative/zero/positive whenever that geodesic is, respectively, timelike/null/spacelike.
The biscalars $u$ and $v$ are  {\it uniquely} determined by the geometry of the spacetime whereas $w$ in principle depends on the quantum state $\ket{\Psi}$.
The term in Eq.~\eqref{eq:Wightman Had} with $u$ is called the {\it direct} part and the term with $v$ the {\it tail} part.

In order to see separately the divergences of the real and imaginary parts of the Wightman function, the following distributional limits are useful:
\begin{align}
&\lim_{\epsilon \rightarrow 0^+}\frac{1}{\sigma \pm \ii \epsilon} =\text{PV}\left({\frac{1}{\sigma}} \right)\mp \ii\pi \delta(\sigma),\\
&\lim_{\epsilon \rightarrow 0^+} \ln\left(\sigma\pm \ii\epsilon\right)=\ln|\sigma|\pm \ii\pi\theta(-\sigma),
\end{align}
where $\text{PV}$ denotes the principal value distribution.
This readily yields the anticommutator
\begin{equation}\begin{split}
&\exptvalstate{\left\{\hat\phi(\coord{x}),\hat\phi(\coord{x}')\right\}}{\Psi}
=
2\text{Re}\left(\Wmanpsix{\coord{x}}{\coord{x}'}\Psi\right)
\\&\quad
=\frac{1}{2\pi^2}
\left[u\, \text{PV}\left(\frac{1}{\sigma}\right)-v\ln\left|\sigma\right|+w\right]
\end{split}\end{equation}
and the commutator 
\begin{equation}\begin{split}
&
\exptvalstate{\left[\hat\phi(\coord{x}),\hat\phi(\coord{x}')\right]}{\Psi}=
2\ii\text{Im}\left(\Wmanpsix{\coord{x}}{\coord{x}'}\Psi\right)
\\&
\quad=\frac{-\sign\left(\Delta t\right)\,\ii}{2\pi}
\left[u\, \delta(\sigma)+ v\,\theta\left(-\sigma)\right)\right].
\end{split}\end{equation}
The anticommutator and commutator (which, when multiplied by `$\ii\, \theta(\Delta t)$' yields the classical retarded Green function) are, respectively dependent and independent of the quantum state $\ket{\Psi}$ of the field.

Eq.~\eqref{eq:Wightman Had} shows explicitly the singularity of the Wightman function along $\sigma=0$, i.e., when $\coord{x}$ and a $\coord{x}'$ in a local neighborhood of $x$ are connected by a null geodesic. 
It is well known~\cite{Garabedian, Ikawa} that the 
Wightman function $\Wmanpsix{\coord{x}}{\coord{x}'}\Psi$ 
continues to diverge
when $\coord{x}$ and $\coord{x}'$ are connected by a null geodesic {\it globally}, i.e., even when $\coord{x}'$ is not in a local neighborhood of $\coord{x}$.
The {\it global} singularity structure of the Wightman function
in the case of Schwarzschild spacetime
was unveiled in Ref.~\cite{BUSS2018168}: 
the divergence of $\Wmanpsix{\coord{x}}{\coord{x}'}\Psi$ 
follows a fourfold pattern, with the singularity type changing every time the null wave front passes through a caustic point (i.e., a spacetime point where neighboring
null geodesics are focused; in Schwarzschild spacetime, because of the spherical symmetry, caustics lie along the line $\gamma=0$  of the point of emission of the null geodesics as well as along the antipodal line $\gamma=\pi$).
Specifically, the pattern for the leading divergence in the real part of the Wightman function is\footnote{An exception to the validity of Eqs.~\eqref{eq:sing Im(GF)} and~\eqref{eq:sing Re(GF)} is at caustic points; also, by ``$\sigma$" in these equations we mean a well-defined extension of the world function outside normal neighborhoods -- see~\cite{casals2016global} about both of these points.}
\begin{equation}
\text{PV}\!\left(\!\frac{1}{\sigma}\!\right)\to-\delta(\sigma)\to-\text{PV}\!\left(\!\frac{1}{\sigma}\!\right)\to\delta(\sigma)\to \text{PV}\!\left(\!\frac{1}{\sigma}\!\right)\to \cdots
\label{eq:sing Im(GF)}
\end{equation}
and that in the imaginary part of the Wightman function is %
\begin{equation}
-\delta(\sigma)\to-\text{PV}\left(\frac{1}{\sigma}\right)\to\delta(\sigma)\to\text{PV}\left(\frac{1}{\sigma}\right)\to-\delta(\sigma)\to\cdots
\label{eq:sing Re(GF)}
\end{equation}
where we have omitted the coefficients of the singularity factors.\footnote{Ref.~\cite{BUSS2018168} showed what the global singularity structure of the {\it Feynman} Green function $G_F\left(\coord{x}',\coord{x}\right)=i \left(\theta(\Delta t)\Wmanpsix{\coord{x}}{\coord{x}'}\Psi+\theta(-\Delta t)\Wmanpsix{\coord{x}'}{\coord{x}}\Psi\right)$ is, 
from which that of the Wightman function  readily follows.}
As an example, the leading singularities of the real and imaginary parts of the Wightman function along the wave front before crossing any caustics are, respectively, $\text{PV}(1/\sigma)$ and $-\delta(\sigma)$ (corresponding to the direct part in Eq.~\eqref{eq:Wightman Had}), whereas after the wave front has crossed one caustic point these turn into, respectively,  $-\delta(\sigma)$ and $-\text{PV}(1/\sigma)$.
This change in the singularity structure will have relevant consequences in entanglement harvesting as we show below.

The above is the leading singularity structure but
there is a corresponding
subleading singularity structure. Ref.~\cite{casals2016global} showed that, for the imaginary part of the Wightman function, its subleading fourfold structure is: 
\begin{equation}\label{eq:sub sing Im}
-\theta(-\sigma)\to\ln\left|\sigma\right|\to \theta(-\sigma)\to -\ln\left|\sigma\right|\to -\theta(-\sigma)\to\cdots
\end{equation}
Ref.~\cite{casals2016global} further conjectured that the subleading structure 
for the real part of the Wightman function is:
\begin{equation}\label{eq:sub sing Re}
-\ln\left|\sigma\right|\to -\theta(-\sigma)\to \ln\left|\sigma\right|\to \theta(-\sigma)\to -\ln\left|\sigma\right|\to\cdots
\end{equation}
The first terms in Eqs.~\eqref{eq:sub sing Im} and~\eqref{eq:sub sing Re} of course correspond to, respectively, the imaginary and real parts of the tail term in Eq.~\eqref{eq:Wightman Had}.

\section{Entanglement harvesting}\label{sec:harvesting}

\subsection{Detector model and perturbative treatment}
Alice and Bob will carry detectors that can locally measure the field around them. To model their detectors we will use the conventional Unruh-DeWitt particle detector model~\cite{dewittQuantumGravityNew1979,Unruh1976}, which consists of a nonrelativistic quantum system coupled locally to a scalar quantum field. 
The Unruh-DeWitt model is covariant and causal\footnote{For pointlike detectors this is strictly true. For smeared detectors this is true in an approximated sense. For details see Refs.~\cite{Martin-MartinezPercheSouza2,deRamonPapageorgiumartin-martinez}}~\cite{martin-martinezCausalityIssuesParticle2015,deRamonPapageorgiumartin-martinez,Martin-MartinezPercheSouza1,Martin-MartinezPercheSouza2} and  captures the main features of the light-matter interaction (e.g., atoms coupled to the electromagnetic field) when exchange of angular momentum between the field 
and the internal degrees of freedom of the detector is not relevant~\cite{Richard}. The covariant treatment and use of time-dependent perturbation theory to calculate the 
joint time evolution of the detectors and the field
in arbitrary curved spacetimes can be found in Refs.~\cite{Martin-MartinezPercheSouza1,Martin-MartinezPercheSouza2}. 
In particular, in the context of entanglement harvesting this model has been extensively used throughout the literature and the particular perturbative approach employed here will use the same notation and conventions used in, among many others, Ref.~\cite{pozas-kerstjensHarvestingCorrelationsQuantum2015}.
Hence, here we only give a brief summary stating the most relevant expressions for the present work.

We model the particle detectors $\dd=\da,\db$ as two-level systems
with energy eigenstates $\ket{g}_\dd$ (ground state) and $\ket{e}_\dd$ (excited state) which are separated by an energy gap $\Omega_\dd$.
The detectors couple to the field amplitude 
$\hat\phi(\coord{x_\dd})$ along their worldline $\coord{x}_\dd(t)$ through the interaction Hamiltonian\footnote{
In the fully covariant formulation of the interaction a Hamiltonian density is prescribed~\cite{Martin-MartinezPercheSouza1}. For pointlike detectors, integrating this density over hypersurfaces of constant $t$ yields Eq.~\eqref{Hamilt1}.}
\begin{equation}\label{Hamilt1}
    \Hit{\dd}t = \lambda_\dd  \eta_\dd(t) \difffrac{\tau_\dd}t \mu_\dd(t) \otimes\hat\phi(\coord{x}_\dd(t)),
\end{equation}
where $\lambda_\dd$ is a coupling constant which is dimensionless in (3+1)-dimensional spacetime,
$0\leq\eta(t)\leq1$ is a real-valued switching function,
$\tau_\dd$ is the detector's proper time
and $\mu_{\dd}(t)= \ee{\ii \Omega_\dd \tau_\dd(t)} \ketbra{e}{g}_\dd+\ee{-\ii \Omega_\dd \tau_\dd(t)} \ketbra{g}{e}_\dd$ 
is the monopole operator. 
Note that the Hamiltonian of~\eqref{Hamilt1} generates time translation with respect to coordinate time $t$, but not the detector's proper time $\tau_\dd$. (For a detailed discussion of this point, see Refs.~\cite{martin-martinezRelativisticQuantumOptics2018,Martin-MartinezPercheSouza1}.)

In the scope of this work, we consider static detectors in Schwarzschild spacetime,  \ie detectors with constant spatial coordinates $\spacoord{x}_\dd=(r_\dd,\theta_\dd,\phi_\dd)$. 
For such detectors
\begin{equation}\label{eq:static proper t}
    \difffrac{\tau_\dd}t = \sqrt{f(r_\dd)}= \sqrt{1-2M/r_\dd},
\end{equation}
and we choose the relation between coordinate time and detector proper time as $\tau_\dd(t)=\sqrt{f(r_\dd)} t$.
As switching functions for the detector we use Gaussians which as a function of coordinate time read 
\begin{equation}\label{eq:Gaussian_switch_function}
    \eta_\dd(t)=\ee{- ((t-t_{0\dd}) /T_\dd)^2},
\end{equation}
where $T_\dd$ denotes the switching width and $t_{0\dd}$ is the center of the switching function.

We assume the initial state of the system (at $t\to-\infty$) to be a product state between the ground states of the two detectors 
and a field state $\rho_\Psi$: 
\begin{equation}\label{eq:rho0}
    \rho_0  =\ketbra{g}g_\da\otimes\ketbra{g}g_\db\otimes\rho_\Psi.
\end{equation}
Assuming that the field state has vanishing one-point function, the expansion of the detectors' state after time evolution $t=0...T$ in coordinate time is
\begin{equation}
\begin{split}
    \rho_{\da\db,T}&= \ketbra{g}g_\da\otimes\ketbra{g}g_\db %
    \\&\quad
    +\lambda_\da^2 \rho_{\da,T}+\lambda_\db^2 \rho_{\db,T}+\lambda_\da\lambda_\db \rho_{\da\db,T}+\mathcal{O}(\lambda^4).
\end{split}
\end{equation}
Using the basis order $\ket{g}_\da\ket{g}_\db, \ket{e}_\da\ket{g}_\db, \ket{g}_\da\ket{e}_\db, \ket{e}_\da\ket{e}_\db $ the final state is represented by the density matrix
\begin{equation}
    \rho_{\da\db,T}=\begin{pmatrix}1-L_{\da\da}^\Psi-L_{\db\db}^\Psi &0&0&(M^\Psi)^*\\0&L_{\da\da}^\Psi&L_{\da\db}^\Psi&0\\ 0&L_{\db\da}^\Psi&L_{\db\db}^\Psi&0\\ M^\Psi&0&0&0
    \end{pmatrix}+\mathcal{O}(\lambda^4),\label{eq:rho_ABT}
\end{equation}
whose entries are 
\begin{widetext}
\begin{align}
    M^\Psi=-\lambda_\da\lambda_\db \integral{t}{-\infty}\infty\integral{t'}{-\infty}t &
    \left( \eta_\da(t) \difffrac{\tau_\da}t \ee{\ii\Omega_\da\tau_\da(t)} \difffrac{\tau_\db}{t'} \eta_\db(t')\ee{\ii\Omega_\db\tau_\db(t')} \Wmanpsix{\coord  x_\da(t)}{\coord x_\db(t')}\Psi \right.
    \nn &\left.\qquad + \eta_\da(t') \difffrac{\tau_\da}{t'} \ee{\ii\Omega_\da\tau_\da(t')} \difffrac{\tau_\db}{t} \eta_\db(t)\ee{\ii\Omega_\db\tau_\db(t)} \Wmanpsix{\coord x_\db(t)}{\coord x_\da(t')}\Psi \right),\label{eq:M_integral_defn}
\\
    L_{\dd\dd'}^\Psi= \lambda_\dd\lambda_{\dd'} \integral{t}{-\infty}\infty\integral{t'}{-\infty}\infty & \eta_\dd(t) \difffrac{\tau_\dd}{t} \eta_{\dd'}(t') \difffrac{\tau_{\dd'}}{t'} \ee{-\ii\Omega_\dd \tau_\dd(t)+\ii\Omega_{\dd'} \tau_{\dd'}(t') } \Wmanpsix{\coord x_\dd(t)}{\coord x_{\dd'}(t')}\Psi , \label{eq:L_integral_defn}
\end{align}
\end{widetext}
where $M^\Psi$ takes complex values while $L_{\dd\dd'}^\Psi\geq0$ takes non-negative values.
For our case of static detectors and Gaussian switching functions, 
as shown in Eqs.~\eqref{eq:L_final_expression} and~\eqref{eq:M_final_expression}, using the mode expansion of the Wightman function, these expressions can be solved analytically up to one integration over the frequency of the field modes which needs to be performed numerically.

To assess and quantify the entanglement of the two detectors in the final state $\rho_{\da\db,T}$ we use its negativity. Its perturbative expansion is %
$\mathcal{N}^\Psi=\max\left[\mathcal{N}^{\Psi,(2)},0\right]+\mathcal{O}(\lambda^4)$ with %
\begin{equation}\label{eq:negativity}
  \mathcal{N}^{\Psi,(2)}=\frac12\left(\sqrt{\left(L_{\da\da}^\Psi-L_{\db\db}^\Psi\right)^2+4\left|M^\Psi\right|^2} -L_{\da\da}^\Psi-L_{\db\db}^\Psi\right).
\end{equation}
We see that whether the two detectors end up in an entangled state, is determined by a competition between the size of the correlating term $M^\Psi$ and the local noise terms $L_{\da\da}^\Psi$ and $L_{\db\db}^\Psi$. In particular, if the noise terms are equal, $L_{\dd\dd}^\Psi:=L_{\da\da}^\Psi=L_{\db\db}^\Psi$, as  will be the case  in Sec.~\ref{sec:results_harvesting}, then the negativity and, thus, the entanglement between the detectors vanishes if $L_{\dd\dd}^\Psi\geq 
\left|M^\Psi\right|$, \ie when the noise overcomes the correlations.
Notice that, because  $L^\Psi_{\da\da}$ and $L_{\db\db}^\Psi$ are local noise terms for which the Wightman function is evaluated along a single detector's worldline, they contain no information about field correlations between the two regions where the two detectors are interacting with the field.

\subsection{When is the entanglement extracted versus generated?}\label{sec:whenisharvestharvest}

Entanglement harvesting is an interesting process because it can demonstrate the presence of entanglement in a quantum field between different spacetime regions. For example, it is clear that when two initially uncorrelated detectors become entangled through their interaction with the field while remaining spacelike separated, then the entanglement they acquire comes from ``extracting" preexisiting entanglement in the field (see, e.g., Refs.~\cite{Reznik2003,pozas-kerstjensHarvestingCorrelationsQuantum2015} in flat spacetime).
However,
when the detectors are not spacelike separated, contributions to the correlating term
$M^\Psi$  in the leading-order perturbative correction to $\rho_{\da\db,T}$ arise which are independent of the quantum state of the field, as recently studied in Ref.~\cite{tjoaWhenEntanglementHarvesting2021}. 
In fact, these contributions can be calculated solely from classical data, consisting of switching functions, detector worldlines and the field classical Green function. Hence, they tell nothing about the quantum properties of the field.

To see this, we follow Ref.~\cite{tjoaWhenEntanglementHarvesting2021}. First, note that the imaginary part of the Wightman function, given by the commutator of the field operators, is independent of the state of the field. Only the real part, which is given by the anticommutator of the field operators, depends on the quantum state.
We can use this to split $M^\Psi$ into two contributions as $M^\Psi=M^{\Psi}_++\ii M^{\Psi}_-$, %
where $M^{\Psi}_+$ is obtained by replacing $W^\Psi$ by its real part (which is symmetric) on the right-hand side of Eq.~\eqref{eq:M_integral_defn}, and $M^{\Psi}_-$ by replacing $W^\Psi$ with its imaginary part (which is antisymmetric). Note that, in general, $M^{\Psi}_+$ and $M^{\Psi}_-$ are complex-valued.

Below we will encounter generic scenarios where the detectors become entangled while $M^\Psi$ is dominated by $M^{\Psi}_-$, and $M^{\Psi}_+$ is (almost or exactly) vanishing.
In such a situation most\footnote{A quantitative study of the relative contributions of $M^{\Psi}_{+}$ and $M^{\Psi}_{-}$ to the entanglement acquired by the detectors for the case of flat spacetime can be found  in Ref.~\cite{tjoaWhenEntanglementHarvesting2021}.} of the entanglement between the detectors is not to be attributed to any preexisting entanglement in the field, as we shall next argue.
In such a scenario the entanglement between the detectors, as measured by $\mathcal{N}^{\Psi,(2)}$, would remain unchanged if we replaced the initial field state by a state which resulted in the same values for $L_{\da\da}^\Psi$ and $L_{\db\db}^\Psi$ (the value of $M^{\Psi}_{-}\approx M^{\Psi}$, would also remain unchanged since it is state independent).
In particular, we could replace the original state of the field by a state that has the same 
Wightman function within the regions where the detectors are coupled to the field, while 
containing no entanglement   between those two regions.

Hence, in this new state, the entanglement between the detectors appears to be generated due to their sequential interaction with the field instead of being extracted from preexisting correlations in the field.
This line of reasoning was used in Ref.~\cite{tjoaWhenEntanglementHarvesting2021} in order to argue that in these cases where $M^{\Psi}_{+}$ is (almost or exactly) vanishing, the entanglement acquired by the detectors should not be referred to as ``entanglement harvesting"  from the field.

In its turn, in a scenario with spacelike-separated detectors where all entanglement between the detectors is harvested from preexisting entanglement in the field, we have that $M^\Psi=M^{\Psi}_{+}$ because the commutator vanishes between the detectors.
That is, in this clear-cut scenario, the contributions from the field's state-dependent anticommutator are the ones that transfer the entanglement from the field to the detectors.

In addition, another observation  made in Ref.~\cite{tjoaWhenEntanglementHarvesting2021} indicates that for timelike-separated detectors the processes captured in $M^{\Psi}_{+}$ are also due to the harvesting of preexisting entanglement in the field 
as opposed to generation of entanglement through the  interaction.
This is based on the fact that a process that creates entanglement between the detectors by having them sequentially interact  with the field  also allows for communication from the first to the second detector: the field carries  information between the two detectors and that can get them entangled independently of any preexisting entanglement in the field. 
However, it is well known that at leading order in perturbation theory, with which we are concerned  here, such causal influence of one detector on the other is determined by the commutator of the field and independent of the anticommutator of the field (see, e.g., Eq.~(24) of Ref.~\cite{ martin-martinezCausalityIssuesParticle2015}). %
Thus, as argued in Ref.~\cite{tjoaWhenEntanglementHarvesting2021}, it appears plausible  that, generally, the detectors are harvesting preexisting entanglement from the field if $M^\Psi\approx M^{\Psi}_{+}$ is dominated by contributions arising from the anticommutator.

\section{Harvesting of gravitationally lensed vacuum entanglement}\label{sec:results_harvesting}

\begin{figure*}
        \centering
        \includegraphics[width = .85\linewidth]{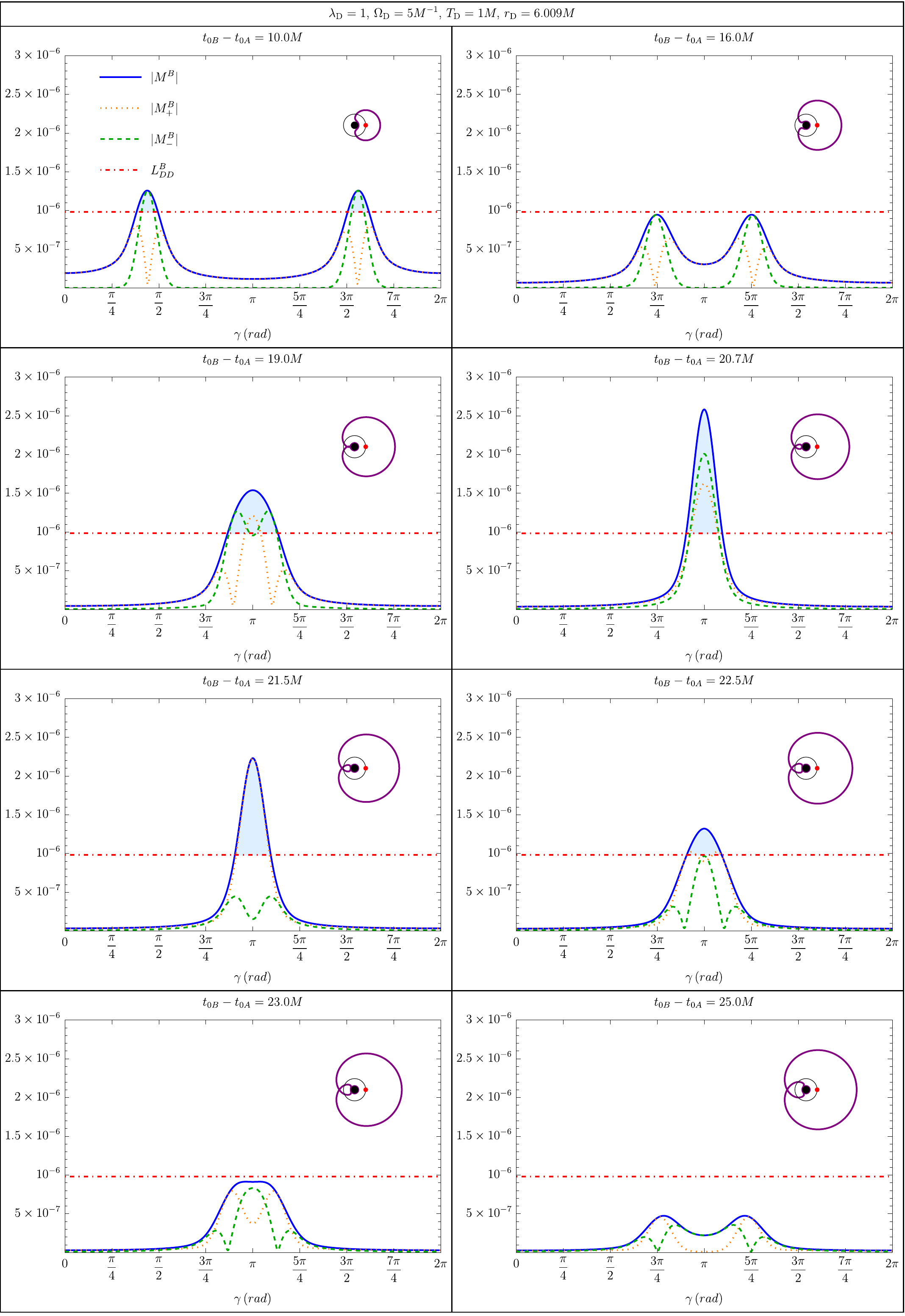}
        \caption{Gravitational lensing of entanglement harvesting from the  Boulware  state: the plots show the  entries $|M^B|$, $|M^B_{\pm}|$ and $|L_{\dd\dd}^B|$ of the final detectors' state~\eqref{eq:rho_ABT} for two static detectors placed at radial coordinate $r=6.009M$ with varying angular separation $\gamma$. 
        Blue shading indicates where $|M^B|>L^B_{\dd\dd}$, \ie the negativity~\eqref{eq:negativity} is positive and the detectors become entangled. All detector parameters are equal ($\lambda_\dd=1,\,\Omega_\dd=5M^{-1},\,T_\dd=1M$); only the delay $\Delta_{\db\da}=t_{0\db}-t_{0\da}$ between the switching functions~\eqref{eq:Gaussian_switch_function} varies between plots. The insets indicate how far a null wave front propagates from the red point within a coordinate time interval $\Delta_{\db\da}$. In particular, at $\Delta_{\db\da}\approx 20.7386M$ at the antipodal point of the red point, \ie at $\gamma=\pi$, the wave front intersects itself and the first caustic point forms. (See supplementary material for an animated version of this plot.)
        }
        \label{plt:nicegrid2by2by}
\end{figure*}

\begin{figure*}
        \centering
        \includegraphics[width = .85\linewidth]{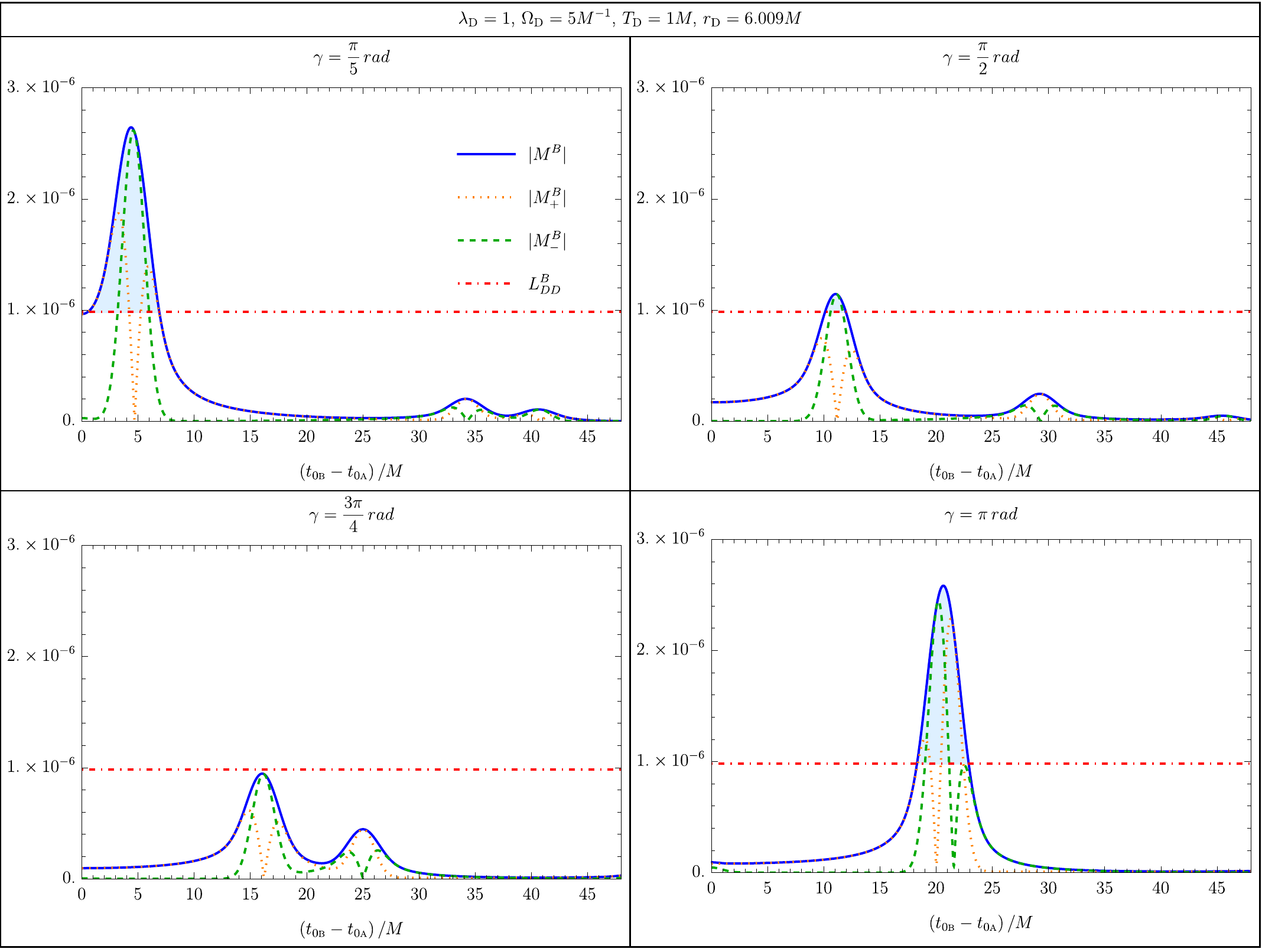}
        \caption{Same data as Fig.~\ref{plt:nicegrid2by2by} from a different perspective: the switching delay $\Delta_{\db\da}=t_{0\db}-t_{0\da}$ is plotted on the horizontal axis, while the panels show detectors with four different angular separations. (See Fig.~\ref{plt:timeonx2by2log} for a logarithmic plot.)
        }
        \label{plt:timeonx2by2}
\end{figure*}

\begin{figure*}
        \centering
        \includegraphics[width = .85\linewidth]{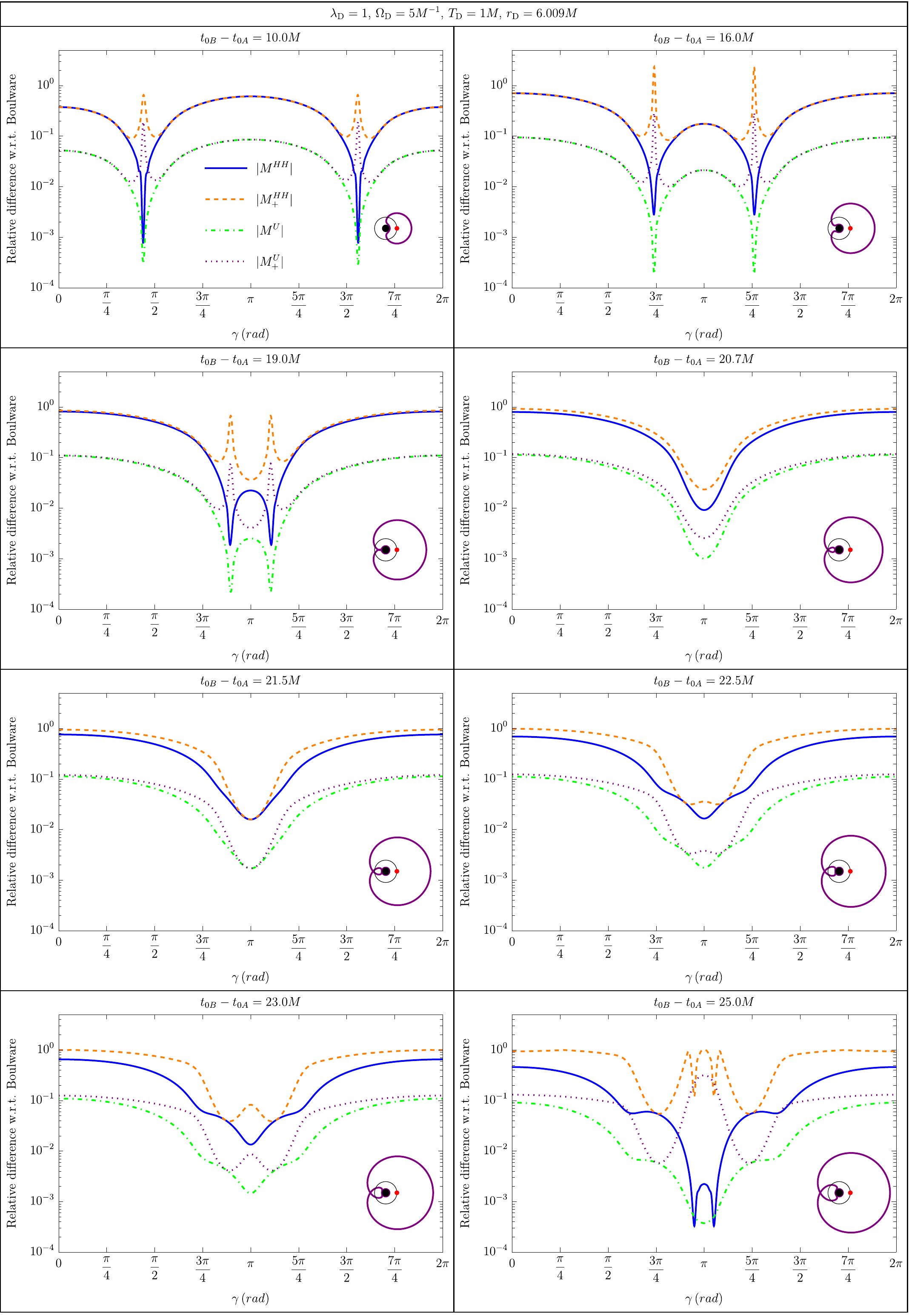}
        \caption{Relative differences between correlation terms in the Boulware state, and the Hartle-Hawking and Unruh states, respectively, \ie plotting   $\frac{|M^H-M^B|}{|M^B|}$, $\frac{|M^U-M^B|}{|M^B|}$, $\frac{|M_+^H-M_+^B|}{|M^B|}$ and $\frac{|M_+^U-M_+^B|}{|M^B|}$, in analogy to Fig.~\ref{plt:nicegrid2by2by}.
        }
        \label{plt:relDifnicegrid2by2by}
\end{figure*}

This section contains the  main results which demonstrate the impact of gravitational lensing, which at caustics refocuses null geodesics emanating from a common source,
on  entanglement harvesting.
To this end, we consider static detectors placed close to the black hole horizon and compare their behavior with the well-studied case of static detectors in flat spacetime.

As one would expect, an important parameter that decides if and to what extent two detectors become entangled is their distance. The distance between  (the static worldlines of) the two detectors can be defined in terms of various meaningful measures (see also Ref.~\cite{jonssonCommunicationQuantumFields2020}). 
In the present context (of static observers in Schwarzschild or Minkowski spacetime), the light propagation coordinate time is an intuitive choice of measure which we use henceforth when referring to the distance between detectors. This is the minimal  amount of (Schwarzschild or Minkowski) coordinate time that it takes light to propagate from the spatial position of one detector to that of the other detector along a null geodesic. 

Once the spatial positions of the two detectors are chosen and, thus, their distance is fixed, the interaction of the detectors with the field can still be made to happen at  spacelike, lightlike or timelike separation by introducing a switching delay; this means that the switching function of the second detector is shifted with respect to the first one by a certain amount of coordinate time. For example, in the following we will assume that  both detectors  couple through the same switching function~\eqref{eq:Gaussian_switch_function} and introduce the switching delay $\Delta_{\db\da}=t_{0\db}-t_{0\da}$ as the (coordinate time) difference between the centers of the switching functions.
The switching delay can then be used to maximize the entanglement in the final state of the two detectors.

In Minkowski spacetime, as far as the impact of the switching delay is concerned, the entanglement in the final state between two detectors at a fixed distance is maximized when the switching functions are exactly null aligned~\cite{tjoaWhenEntanglementHarvesting2021}, \ie when the switching delay equals the light propagation time.\footnote{This statement is not entirely precise, as in Minkowski spacetime the maximum is located very close to--but not exactly at--null alignment of the switching functions, see Ref.~\cite{tjoaWhenEntanglementHarvesting2021}.}
As far as the impact of the distance between the detectors is concerned, its impact on the final entanglement is straightforward: if the distance between the detectors is increased (while allowing the switching delay to be optimal and leaving other coupling parameters unchanged), the entanglement between the detectors in their final state decreases and goes to zero beyond a certain distance.
This holds even if the switching delay is readjusted to uphold null alignment.
Furthermore, in 3+1-dimensional Minkowski spacetime, the entanglement between exactly null aligned detectors is dominated by correlations generated by communication rather than by harvesting, because the $M^\Psi$ term is dominated by $M^\Psi_-$ rather than $M^\Psi_+$, as shown in Ref.~\cite{tjoaWhenEntanglementHarvesting2021}.

This raises two questions regarding  entanglement harvesting in Schwarzschild spacetime.
First, does the presence of caustics enhance the ability of detectors to become entangled?
What is more, can caustics make the harvested entanglement  no longer decrease  monotonically with distance? 

Second, does the singularity structure of the Wightman function allow for the entanglement to be dominated by harvesting rather than signaling if the detectors' switching instead of being aligned along a direct null geodesic is aligned along a secondary null geodesic, \ie a null geodesic that has passed through a caustic? As discussed in Sec.~\ref{sec:Wman_sing_structure}, here the singularity structures of the real and imaginary parts of the Wightman function are shifted, so that  $M^\Psi_+$ could dominate over $M^\Psi_-$ because  the real part now carries the $\delta(\sigma)$-singularity.
This  would then constitute entanglement harvesting between timelike-separated detectors, since the secondary null geodesics lie inside the causal cone which is bounded by the direct null geodesics between detectors.

The following results answer both questions in the affirmative.

Fig.~\ref{plt:nicegrid2by2by} shows the gravitational lensing effect on the harvesting of entanglement by two static detectors in Schwarzschild spacetime when the field is in the Boulware state. The scenario is as follows.
Two identical detectors, with detector gap $\Omega_{\da\da}=\Omega_{\db\db}=5M^{-1}$ and $\lambda=\lambda_\da=\lambda_\db$,\footnote{Note that because $\lambda$ is dimensionless and all terms are of order $\mathcal{O}(\lambda^2)$ the numeric value of $\lambda$ only affects the result by an overall factor. We choose $\lambda=1$ for improved readability.}
are placed on static worldlines at the same radial coordinate $r=6.009M$. The angular separation $\gamma$ between the two detectors varies along the horizontal axes of each panel.
Both detectors are coupled to the field through the Gaussian switching function~\eqref{eq:Gaussian_switch_function}, but  the difference between  the two centers of the switching functions  is shifted by an amount of coordinate time $\Delta_{\db\da}=t_{0\db}-t_{0\da}$   which varies between the panels as indicated.

An inset in each panel visualizes how far null geodesics, emanating from detector $\da$ at time $t_{0\da}$, propagate within the coordinate time interval $\Delta_{\db\da}$:
the black filled circle represents the interior of the black hole horizon at $r=2M$,  a black ring around it represents all points with radial coordinate $r=6.009M$ and the red dot represents the spatial location of detector $\da$. 
The violet curve then indicates the  spatial  position of the null wave front after propagation time $\Delta_{\db\da}$.
In particular, the different insets show the formation of  caustics located at angular separation $\gamma=\pi$ from the point of origin (which is plotted in red) where the null wave front intersects itself. The caustic  reaches $r=6.009M$ at $\Delta_{\db\da}\approx 20.7386M$.

The four different curves in each panel show the resulting contributions to the final density matrix of the two detectors~\eqref{eq:rho_ABT}, when the field is in the Boulware state. (Accordingly they are denoted with a superscript $\Psi=B$.)
The horizontal, red and dashed line shows the single detector noise term $L^B_{\dd\dd}:=L^B_{\da\da}=L^B_{\db\db}$. Because it only depends on the radial coordinate and the switching function's width $T_{\dd}$, it is equal for both detectors and constant for all positions of detectors considered here.%
The blue solid curve shows the absolute value $|M^B|$ resulting from the chosen value of the switching delay $\Delta_{\db\da}$ in each panel, as a function of the angular separation between the detectors.
The two remaining dashed curves show the absolute values $|M_-^B|$ 
in green and $|M_+^B|$ in orange which, as discussed in Sec.~\ref{sec:whenisharvestharvest}, are obtained by replacing the Wightman function by its imaginary and real parts, respectively. 
In particular, as discussed in Sec.~\ref{sec:whenisharvestharvest}, if $|M^B|$ is dominated by $|M^B_-|$ (obtained from the imaginary part), then the detectors' entanglement is predominantly generated in the sequential interaction of the detectors with the field. On the other hand, if $|M^B|$ is dominated by $|M^B_+|$ then final entanglement between the detectors is predominantly harvested from preexisting entanglement in the vacuum of the field.

Altogether the panels in  Fig.~\ref{plt:nicegrid2by2by} illustrate the lensing effect which appears in proximity to the caustics.
The first two panels show the settings with the shortest switching delay $\Delta_{\db\da}$ between the switching functions.
In these two plots, we see that the correlations between the detectors, as captured by $|M^B|$, are largest when the centers of the intervals during which the detectors couple to the field are connected by (direct) null geodesics.
In the first panel there is a certain interval of angular separations in which the blue line for $|M^B|$ exceeds the noise term $L^B_\dd$
and the detectors end up entangled.
However, in the second panel  the detectors' final state remains separable, even for null separated detectors, because $\Delta_{\db\da}$ is increased.
Note that here the contribution from $M^B_-$ dominates the peaks of $M^B$; hence entanglement between the detectors cannot be attributed to harvesting of preexisting entanglement from the field but to entanglement generation due to sequential interaction. 

In flat spacetime  no entanglement would be observed in the final detectors' state for larger switching delay between the detectors.
Intuitively this is related to the growth of the light cone's surface area which dilutes the entanglement.
In Schwarzschild spacetime, however, where the light cone refocuses at caustics the opposite can happen. As seen in the third  through sixth panels  (in order of increasing $\Delta_{\db\da}$), the detectors can become entangled at larger switching delays $\Delta_{\db\da}$ again, at angular separations close to $\gamma=\pi$ in the proximity of caustics.
This answers  in the affirmative the first of the two guiding questions raised above.

The second of the two questions raised above pertains to the Wightman function's singularity structure  described in Sec.~\ref{sec:Wman_sing_structure}. Its effect is easy to recognize when comparing the last (bottom right) panel to the first (top left) panel of Fig.~\ref{plt:nicegrid2by2by}.
In both panels the correlations between the two detectors, as captured by $|M^B|$, are maximal for null aligned detectors. However, in the first panel the detectors are connected by primary null geodesics, whereas in the last panel they are connected by secondary null geodesics. At secondary null geodesics, as discussed in Sec.~\ref{sec:Wman_sing_structure}, the singularity structure of the Wightman function is shifted from primary null geodesics so that now the real (anticommutator) part carries the $\delta(\sigma)$-singularity, which used to be carried instead by the imaginary (commutator) part in the case of  primary null geodesics.
Accordingly, the two contributions $|M^B_\pm|$ have a qualitatively similar overall shape in the first and the last panels, except that the $M^B_+$ and $M^B_-$ have swapped places. For detectors aligned along a secondary null geodesic, the real part contribution $M^B_+$ dominates over the imaginary part contribution $M^B_-$.
So if these correlations overcame the noise, $|M^B|>|L_{\dd\dd}|$, the detectors would become entangled by harvested entanglement.
However, the correlations in the last panel are too weak for the detectors to end up entangled. To find detectors that get entangled by genuinely harvested correlations around secondary null geodesics, we need to position the detectors closer to caustics so as to make use of the overall enhancement of the Wightman function there.
Such a setting is seen in the fifth panel (for $\Delta_{\db\da}=21.5M$) of Fig.~\ref{plt:nicegrid2by2by}. 
In this and the following panels
of Fig.~\ref{plt:nicegrid2by2by},
the switching delay $\Delta_{\db\da}$ is larger than the shortest light propagation (Schwarzschild coordinate) time between detectors with angular separation $\gamma=\pi$, and 
the detectors (at $r=6.009M$) which are null aligned here are aligned along a secondary null geodesic and timelike separated.
With this we have answered  in the affirmative the second of the above questions, as we observe the genuine harvesting of preexisting entanglement from the field by timelike-separated detectors, aligned along secondary null geodesics. In the following, we will see further examples allowing for entanglement harvesting between timelike-separated detectors with angular separation $\gamma=\pi$.

Complementary to Fig.~\ref{plt:nicegrid2by2by}, the same scenario and effects are seen in Fig.~\ref{plt:timeonx2by2} from a slightly different perspective. Here, 
the contributions to the detectors' final density matrix are plotted over the coordinate time switching delay $\Delta_{\db\da}$, while the four different panels correspond to four different angular separations. (Fig.~\ref{plt:timeonx2by2log} provides a logarithmic plot of the same data.)
Note that these plots over $\Delta_{\db\da}$ are directly comparable to the plots also found in Ref.~\cite{tjoaWhenEntanglementHarvesting2021} in 3+1-dimensional Minkowski spacetime.

In the first panel of Fig.~\ref{plt:timeonx2by2}, for angular separation $\gamma=\pi/5$, there are three peaks appearing in $|M^B|$ which correspond to an alignment of the switching functions along primary, secondary and tertiary null geodesics, respectively.
The qualitative structure of the first peak and its contributions from $|M_+^B|$ and $|M_-^B|$ correspond to the structure found in flat spacetime in Ref.~\cite{tjoaWhenEntanglementHarvesting2021}. However, in the secondary peak the qualitative roles of $|M_+^B|$ and $|M_-^B|$ are interchanged due to the shifted singularity structure of the Wightman function, as is easy to see in the logarithmic plot in Fig.~\ref{plt:timeonx2by2log}. For the tertiary peak, together with the singularity structure of the Wightman function, the qualitative structure of the peak also shifts back to its primary form.

Even if there are three peaks appearing at angular separation $\gamma=\pi/5$, only the first one corresponding to alignment along the direct, primary null geodesic exhibits entanglement in the  detectors' final state.
In an intermediate regime of angular separations, for $0.64\pi
\lesssim \gamma \lesssim 0.81\pi$,  not even the primary peak overcomes the noise and the detectors' final state remains separable for all switching delays $\Delta_{\db\da}$.
An example of this is seen in the third (bottom left) panel of Fig.~\ref{plt:timeonx2by2} for angular separation $\gamma=3\pi/4$.
However, as the angular separation approaches $\gamma=\pi$ both the primary and secondary peaks increase their size again and they overcome the noise, thus, leaving the detectors in an entangled state.
Gradually, as $\gamma\to\pi$, the primary and secondary peaks superpose and finally create one joint peak aligned at the caustic for $\gamma=\pi$. Here we see that for switching delays that are somewhat larger than the direct null alignment, the extracted entanglement can be dominated by $|M_+^B|$ and thus can be attributed to entanglement harvesting from the field.

\begin{table}
    \begin{tabular}{||l|l||}
    \hline
    \textbf{State} & \textbf{Local noise term} \\
    \hline
    Boulware state   & $L^{B}_{DD}/(\lambda_\da\lambda_\db) =9.82\times 10^{-7}$  \\ 
    Unruh state & $L^{U}_{DD}/(\lambda_\da\lambda_\db) =9.96\times 10^{-7}$  \\ 
    Hartle-Hawking state & $L^{H}_{DD}/(\lambda_\da\lambda_\db) = 1.08\times 10^{-6}$ \\ \hline
    \end{tabular}\caption{
    Numerical values of the local noise terms in the three different states considered, for the scenarios in Fig.~\ref{plt:nicegrid2by2by}, Fig.~\ref{plt:nicegrid2by2byU} and Fig.~\ref{plt:nicegrid2by2byH}, respectively.
    }
    \label{tbl: Noise terms}
\end{table}

The results shown in Fig.~\ref{plt:nicegrid2by2by} are for the field in the Boulware state. As seen in the analogous Fig.~\ref{plt:nicegrid2by2byU} for the Unruh state 
and Fig.~\ref{plt:nicegrid2by2byH} 
for the Hartle-Hawking state,
the same phenomena appear in these states as well.
In fact, the quantitative differences between the states (which may be difficult to see with the naked eye) are mostly due to the difference in the noise term, which is given in Tab.~\ref{tbl: Noise terms}. The correlation terms, in particular in the scenarios of detectors that harvest entanglement in the proximity of caustics, agree to many digits for all three states, as can be seen in Fig.~\ref{plt:relDifnicegrid2by2by}, which shows their relative differences.

\begin{figure*}
        \centering
        \includegraphics[width = .85\linewidth]{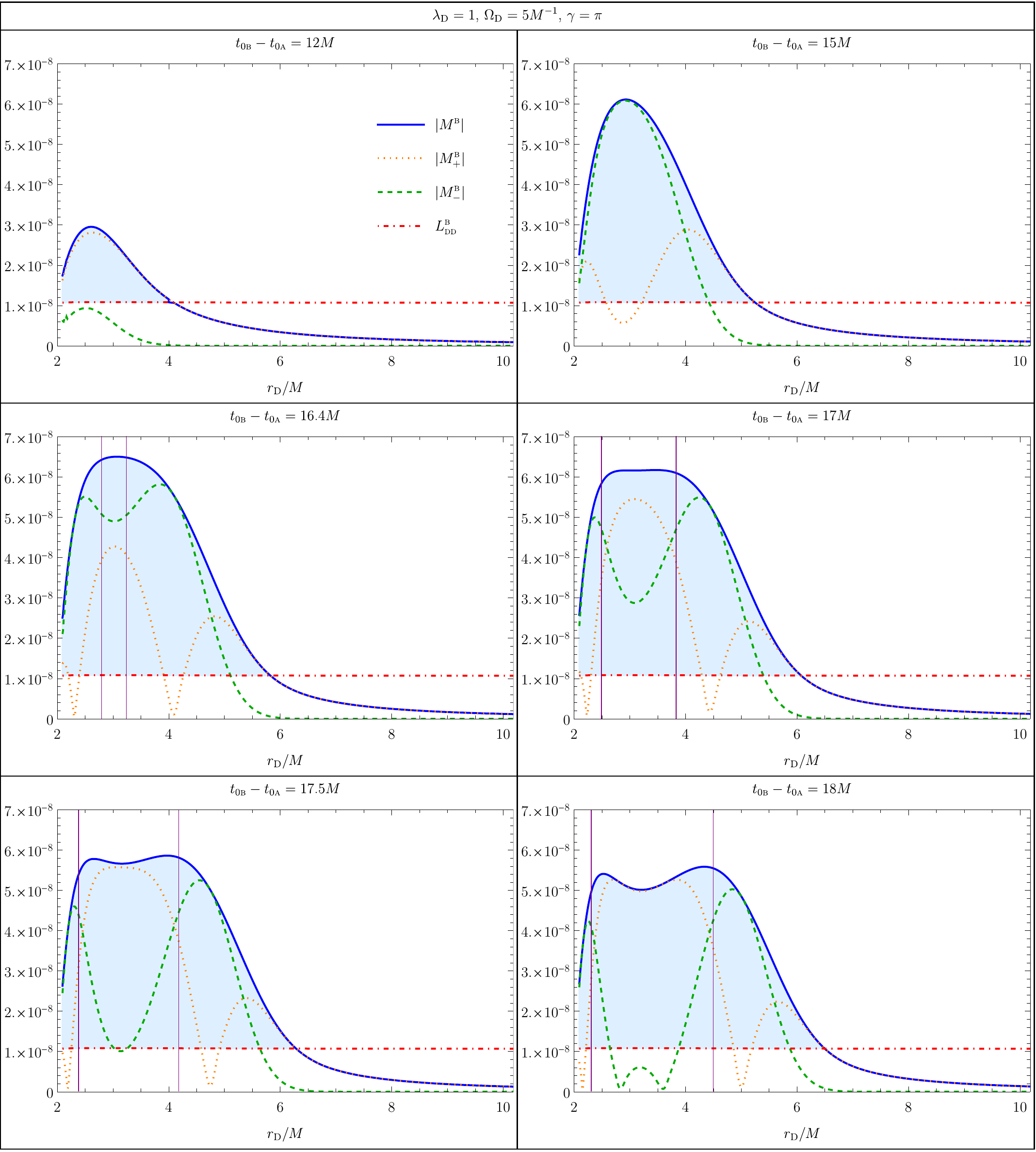}
        \caption{Entanglement harvesting from the Boulware state by static detectors at antipodal positions close to a Schwarzschild black hole: the plots show the entries of the final detectors' state~\eqref{eq:rho_ABT} for two identical detectors ($\lambda_\dd=1, \Omega_\dd=5/M$) which are placed at equal radial coordinates $r_\dd$ at angular separation $\gamma=\pi$, \ie with the black hole exactly in the middle between them. 
        To account for the different redshifts the width of the detector switching functions~\eqref{eq:Gaussian_switch_function} is set to  $T_\dd= M/\sqrt{1-2M/r}$, \ie kept equal with respect to the detectors' proper time.
        The different panels show the results for different switching delays $\Delta_{\db\da}=t_{0\db}-t_{0\da}$. Detectors placed at $r=3M$ are the first for which the switching functions are exactly null aligned at $\Delta_{\db\da}=3\sqrt3\pi M\approx 16.32M$. Both for larger and lower radial coordinates the light propagation time is longer, and hence two peaks form in the later panels around the radial coordinates for which exact null alignment is achieved (see Fig.~\ref{plt:propagation_times_antipodal}). The vertical purple lines indicate the radial coordinates at which null alignment of the switching function is achieved, \ie where $\Delta_{\db\da}=\Delta t$ (see Fig.~\ref{plt:propagation_times_antipodal}). (See supplementary material for animated version of this plot.)
        }
        \label{plt:negativity_antipodal_detectors}
\end{figure*}
\begin{figure}
        \centering
        \includegraphics[width = .95\columnwidth]{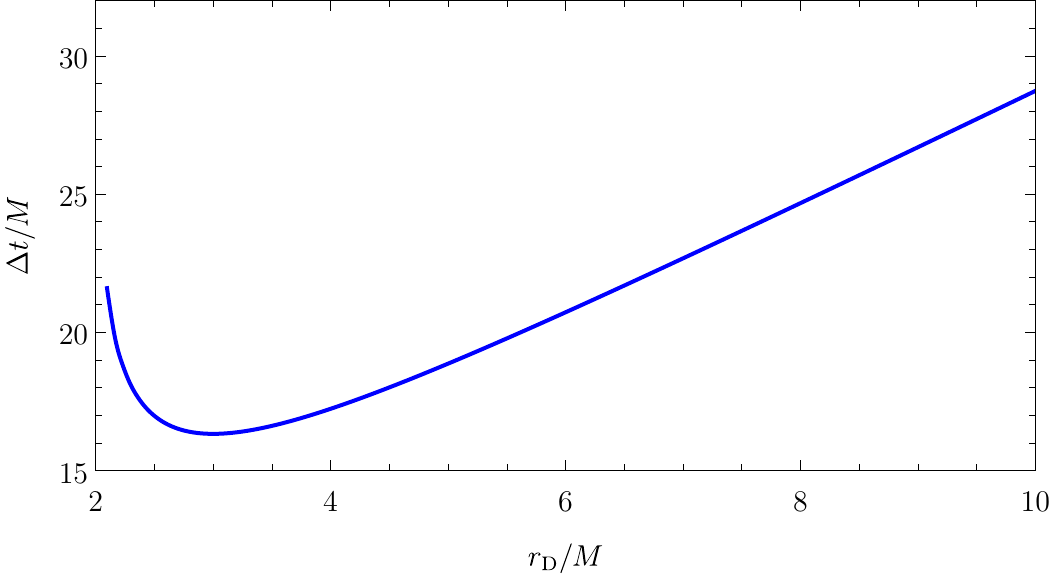}
        \caption{Light propagation coordinate time between detectors in the scenario of Fig.~\ref{plt:negativity_antipodal_detectors}.
        }
        \label{plt:propagation_times_antipodal}
\end{figure}

We conclude this section with a study of entanglement harvesting from the Boulware state between detectors at antipodal locations with the black hole exactly in the middle between them, i.e., at identical radial coordinates $r=r_\da=r_\da$ and with an angular separation $\gamma=\pi$, for a range of detector locations reaching down very close to the horizon at radial positions $r=2.095M$.

At the different radial coordinates the detectors experience different gravitational redshifts. To account for this, we adjust the width of the switching functions~\eqref{eq:Gaussian_switch_function} to $T_\dd(r)=T_{\infty}/\sqrt{f(r)}=T_\infty/\sqrt{1-2M/r}$, such that, at all radial coordinates $T_\dd(r)$, corresponds to the same amount of proper time given by some value $T_\infty$.
Analogous to the plots above, Fig.~\ref{plt:negativity_antipodal_detectors} %
shows the resulting contributions to the detectors' final state for  $T_{\infty}=1M$.
It shows that at all radial coordinates considered, even close to the horizon, the two detectors can become entangled and the entanglement can be dominated by entanglement harvesting, when the switching delay $\Delta_{\db\da}=t_{0\db}-t_{0\da}$ is chosen appropriately.

To understand the structure of the plots and its dependence on $\Delta_{\db\da}$ it is useful to consider the  light propagation coordinate time $\Delta t$ between the two detectors. This is the coordinate time that it takes for a null geodesic starting at spatial coordinates $(r,\theta=\pi/2,\phi)$ to reach the antipodal point $(r,\theta=\pi/2,\phi+\pi)$.
Fig.~\ref{plt:propagation_times_antipodal} shows this time for the range of radial coordinates we consider.  It shows that $\Delta t$ is minimal at $r=3M$. 
For this critical case, it takes the value $\Delta t_{\text{min}}=3\sqrt{3}\pi M$.
Setting the switching delay equal to the light propagation time, \ie setting $\Delta_{\db\da}=\Delta t$, yields exact null alignment of the detectors' switching functions. 

Around exact null alignment between the detectors we expect a peak in the final detector entanglement, and this is, indeed, what we observe in Fig.~\ref{plt:negativity_antipodal_detectors}. At first, in the panels showing the lower values of $\Delta_{\db\da}$, one peak forms around $r=3M$ since the detectors located at this radial coordinate are the first to be exactly null aligned.
As $\Delta_{\db\da}$ is increased in the following panels, a double-peak structure forms since for $\Delta_{\db\da}>3\sqrt{3}\pi M$ there are always two radial coordinate positions at which the detectors are exactly null aligned.
Regarding the relative size of the contributions $M_+^B$  and $M_-^B$, we observe again that both contributions are of comparable size when the detectors are exactly null aligned, as we already observed in the panel of Fig.~\ref{plt:timeonx2by2}, for $\gamma=\pi$.
This is in contrast to scenarios where the detectors are null aligned along primary null geodesics and sufficiently far away from any caustics. There, just as in flat Minkowski spacetime, the correlations are dominated by the $M_-$ contribution when the detectors are exactly null aligned. However, at the caustics, where a whole envelope of null geodesics connects the detectors at once, the singularity structure of the two-point function is altered (see Ref.~\cite{casals2016global}). This results in the $M_+$ contribution dominating before exact null alignment and the $M_-$ contribution dominating after exact null alignment of the switching functions.

\section{Discussion and outlook}\label{sec:outlook}

Using UDW detectors, we investigated entanglement harvesting from a Klein-Gordon field in the Boulware, Hartle-Hawking and Unruh vacuum states in the background of Schwarzschild spacetime in 3+1 dimensions. We showed that the realistic 3+1-dimensional case possesses a particularly rich phenomenology due to the presence of secondary (and higher) null geodesics and caustics. 

In particular, we investigated the ability of two detectors that are static at the same radial coordinate, but with different angular coordinates, to harvest entanglement from the Boulware, Hartle-Hawking and Unruh vacua as they get closer to the horizon.

We paid special attention to entanglement harvesting in the regions of spacetime where secondary null geodesics and caustics  can connect the two harvesting detectors. 
We found that genuine entanglement harvesting can indeed be amplified through ``entanglement gravitational lensing" effects when the detectors  
are close to regions where caustics appear. By the term ``genuine entanglement harvesting" we mean the extraction of preexisting entanglement from the field, as opposed to the extraction of entanglement that is created through communication between the detectors.
Interestingly, this genuine harvesting can also appear for timelike-separated detectors, for example, when the detectors are connected by secondary null geodesics.
Mathematically, this results from a change in the singularity structure of the Wightman function as the field waves cross through caustics.

The formalism that we developed here can also be used to analyze further interesting black hole entanglement harvesting configurations, such as the case where the detectors are in motion and possibly also when crossing the event horizon. 

Furthermore, since the effects of entanglement harvesting amplification that we found here  are due to lensing which arises from the existence of secondary null geodesics connecting the two detectors, an intriguing follow-up question arises: to what extent could entanglement harvesting be engineered to be amplified even in flat spacetime, namely in the presence of suitable mirror and lensing arrangements?  
These scenarios will be explored in future work.

\section*{Acknowledgments}
J.G.A.C. acknowledges financial support by CNPq (Brazil), Grant number 1424412/2019-8.
R.H.J. gratefully acknowledges support by the Wenner-Gren Foundations and, in part, by the Wallenberg Initiative on Networks and Quantum Information (WINQ).
Nordita is supported in part by NordForsk.
M.C. acknowledges partial financial support by the Scientific Council of the Paris Observatory during a visit. 
E.M.M. acknowledges support through the Discovery Grant Program of the Natural Sciences and Engineering Research Council of Canada (NSERC). EMM also acknowledges support of his Ontario Early Researcher award.
A.K. acknowledges support through a Discovery Grant from the National Science and Engineering Research Council of Canada (NSERC) and a Discovery Project Grant from the Australian Research Council (ARC).

\appendix
\begin{figure*}
        \centering
        \includegraphics[width = .85\linewidth]{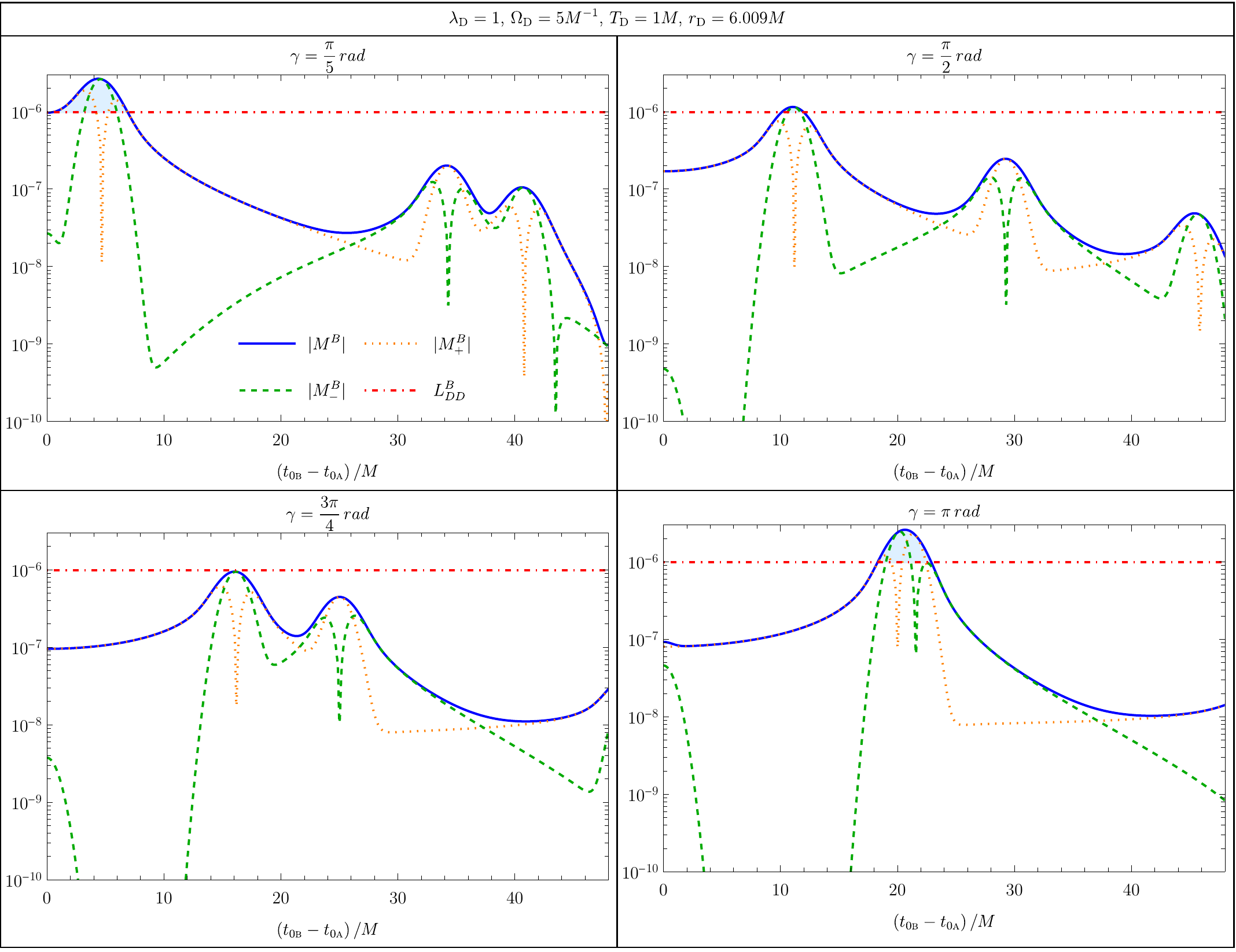}
        \caption{Logarithmic plot version of Fig.~\ref{plt:timeonx2by2}. (See supplementary material for animated version of this plot.)
        }
        \label{plt:timeonx2by2log}
\end{figure*}

\begin{figure*}
        \centering
        \includegraphics[width = .85\linewidth]{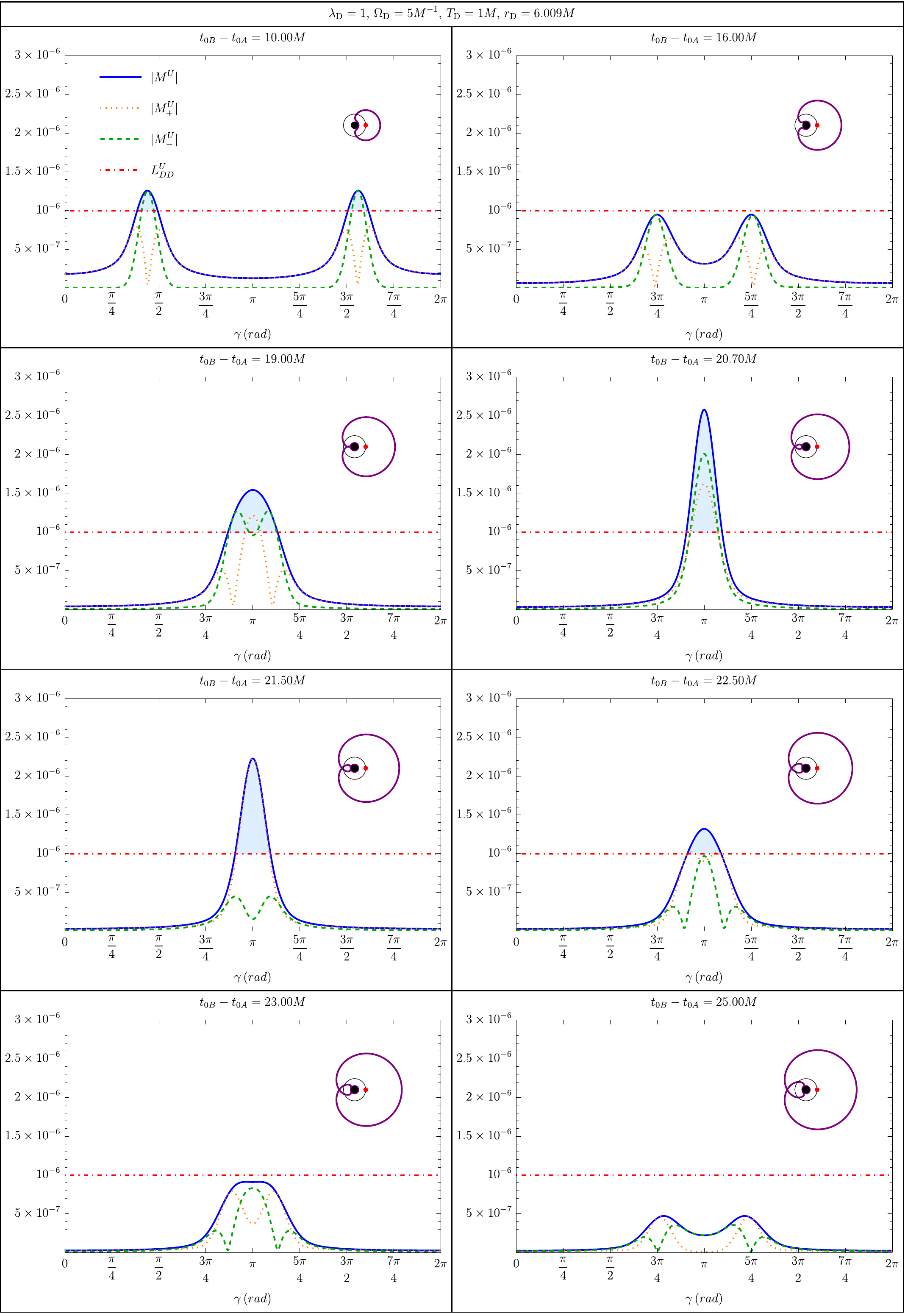}
        \caption{
        Gravitational lensing of entanglement harvesting  from the field in the Unruh state: 
        the entire setup is identical to the setup of Fig.~\ref{plt:nicegrid2by2by}, except that here the initial state for the field is the Unruh state.
        The plots  show the absolute values of $M^U$, $M^U_{
        \pm}$ and $L_{\dd\dd}^U$ in Eq.~\eqref{eq:rho_ABT} for two static detectors placed at radial coordinate $r=6.009M$ with varying angular separation $\gamma$. All detector parameters are equal ($\lambda_\dd=1,\,\Omega_\dd=5M^{-1},\,T_\dd=1M$); only the offset $\Delta_{\db\da}=t_{0\db}-t_{0\da}$ between the two switching functions~\eqref{eq:Gaussian_switch_function} varies between the plots. The inset indicates how far a null wave front propagates from the red point within a coordinate time interval $\Delta_{\db\da}$.
        }
        \label{plt:nicegrid2by2byU}
\end{figure*}

\begin{figure*}
        \centering
        \includegraphics[width = .85\linewidth]{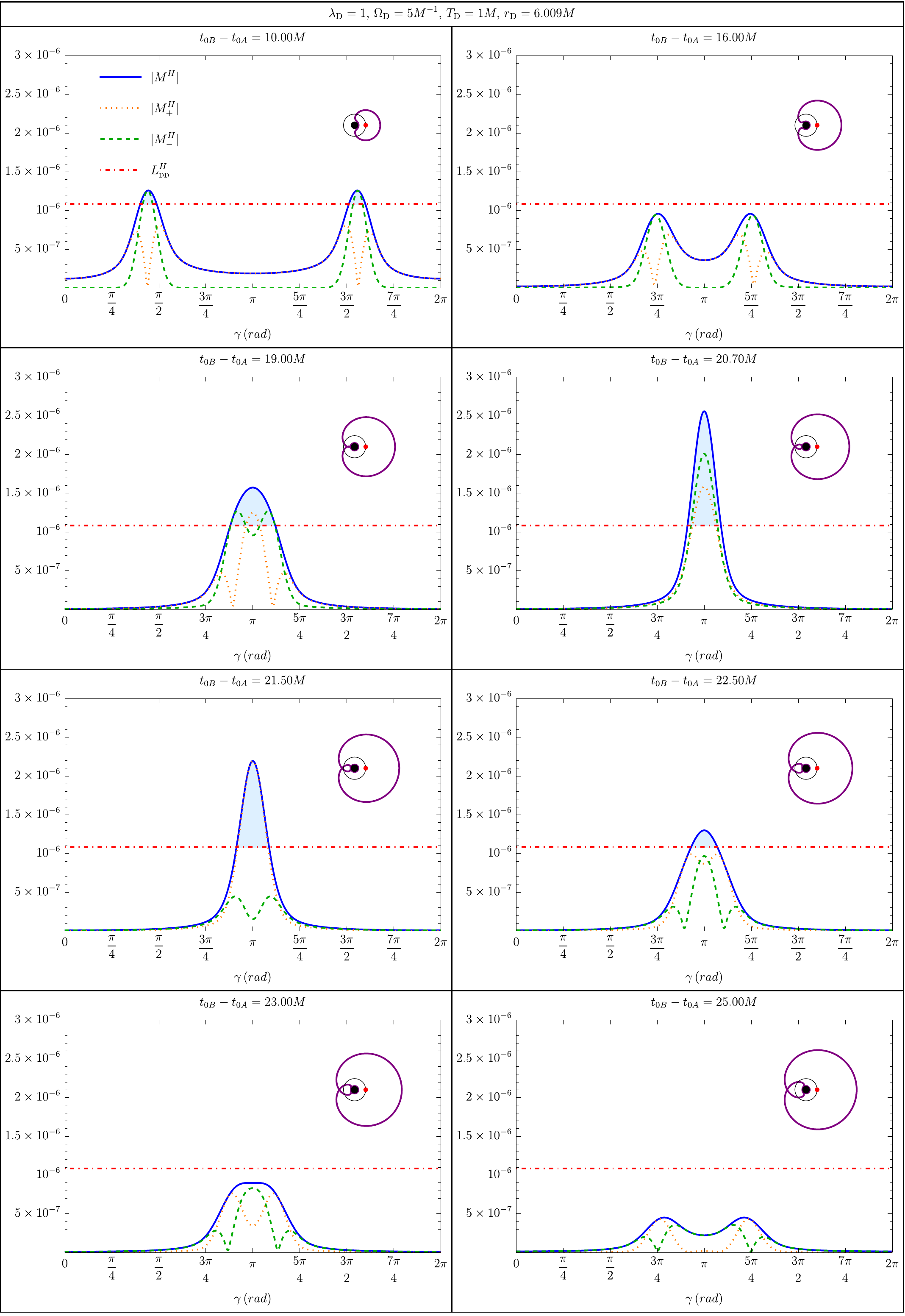}
        \caption{Gravitational lensing of entanglement harvesting from the field in the Hartle-Hawking state: 
        the entire setup is identical to the setup of Fig.~\ref{plt:nicegrid2by2by}, except that here the initial state for the field is the Hartle-Hawking state.
        The plots  show the absolute values of $M^H$, $M^H_{\pm}$ and $L_{\dd\dd}^H$ in Eq.~\eqref{eq:rho_ABT} for two static detectors placed at radial coordinate $r=6.009M$ with varying angular separation $\gamma$. All detector parameters are equal ($\lambda_\dd=1,\,\Omega_\dd=5M^{-1},\,T_\dd=1M$); only the offset $\Delta_{\db\da}=t_{0\db}-t_{0\da}$ between the two switching functions~\eqref{eq:Gaussian_switch_function} varies between the plots. The inset indicates how far a null wave front propagates from the red point within a coordinate time interval $\Delta_{\db\da}$.
        }
        \label{plt:nicegrid2by2byH}
\end{figure*}

\begin{widetext}
\section{Density matrix contributions}\label{app:LM_expressions}
In this appendix, we give a detailed derivation of the expressions used for the calculation of the perturbative contributions to the final density state of the two detectors in Eq.~\eqref{eq:rho_ABT}.
To evaluate the leading-order contributions to the detectors' final density matrix, we need certain Fourier-type integrals of the switching function~\eqref{eq:Gaussian_switch_function},
$\eta_\dd(t)=\ee{- ((t-t_{0\dd}) /T_\dd)^2}$.
First, we have
\begin{align}\label{eq:1st int Gauss}
&\integral{t}{-\infty}s \ee{\ii \nu t} \eta_\dd(t)
 = \frac{\sqrt\pi T_\dd}2 \ee{-\frac{\nu^2 T_\dd^2}4} \ee{\ii\nu t_0\dd} \left(1+\text{erf}\left(-\frac{\ii\nu T_\dd}2 +\frac{s-t_{0\dd}}{T_\dd}\right)\right)\text{,}
   \\ \label{eq:2nd int Gauss}
&   \integral{t}{-\infty}\infty \ee{\ii \nu t} \eta_\dd(t) = \sqrt{\pi } T_\dd \ee{-\frac{1}{4} \nu^2 T_\dd^2} \ee{\ii\nu t_{0\dd}}\text{,}
\end{align}
where $\text{erf}(z)=\frac{z}{\sqrt\pi} \integral{t}0z\ee{-t^2}$.
Furthermore, we need the integral
\begin{equation}\label{eq:Double time integral of the switching}
    \integral{t'}{-\infty}\infty \ee{\ii\mu t'} \eta_{\dd'}(t')\integral{t}{-\infty}{t'}\ee{\ii\nu t}\eta_\dd(t).
\end{equation}
To this end, consider
\begin{equation}
\begin{split}
      &\frac{\sqrt\pi T_\dd}2 \ee{-\frac{\nu^2 T_\dd^2}4} \ee{\ii \nu t_0\dd}  
      \integral{t'}{-\infty}\infty \ee{\ii\mu t'} \ee{-\frac{(t'-t_{0\dd'})^2}{T'^2_\dd}} 
       \left(1+\text{erf}\left(-\frac{\ii\nu T_\dd}2 +\frac{t'-t_{0\dd}}{T_\dd}\right)\right)
     \\&
     =\frac{\sqrt\pi T_\dd}2 \ee{-\frac{\nu^2 T_\dd^2}4} \ee{\ii\nu t_0\dd}  
       \ee{-\frac{t_{0\dd'}^2}{T_{\dd'}^2} } \integral{t'}{-\infty}\infty \ee{-\frac{t'^2}{T_{\dd'}^2} +t'\left( 2\frac{ t_{0\dd'}}{T_{\dd'}^2}+\ii\mu  \right)}
       \text{erf}\left(-\frac{\ii\nu T_\dd}2 +\frac{t'-t_{0\dd}}{T_\dd}\right) 
       +  \frac{ \pi T_\dd T_{\dd'}}2 \ee{-\frac{\mu^2 T_{\dd'}^2+\nu^2 T_\dd^2}4} \ee{\ii(\mu t_{0\dd'}+\nu t_0\dd)}.   
\end{split}
\end{equation}
For the $t'$-integral, use Eqs.~(A3) and~(A7) from Ref.~\cite{pozas-kerstjensHarvestingCorrelationsQuantum2015}, which  is
\begin{equation}
    I(a,b)= \integral{y}{-\infty}\infty \ee{-a^2-\ii b y -y^2}\text{erf}(y-\ii a)= -\ii\sqrt{\pi} \ee{-a^2-b^2/4}\text{erfi}\left(\frac{a+b/2}{\sqrt2}\right)
,
\end{equation}
where $\text{erfi}(z)=-\ii\,\text{erf}(\ii z)$,
with $y=t'/T_{\dd'}$, $a=\frac{\nu T_\dd}2-\ii\frac{t_{0\dd}}{T_\dd}$
and  $b= 2\ii\frac{ t_{0\dd'}}{T_{\dd'}}-\mu T_{\dd'} $, then
\begin{equation}
\begin{split}
    & 
    \integral{t'}{-\infty}\infty \ee{-\frac{t'^2}{T_{\dd'}^2} +t'\left( 2\frac{ t_{0\dd'}}{T_{\dd'}^2}+\ii\mu  \right)}
       \text{erf}\left(-\frac{\ii\nu T_\dd}2 +\frac{t'-t_{0\dd}}{T_\dd}\right) 
    \\& 
    =   -\ii\sqrt\pi T_{\dd'} \ee{\frac{-\mu^2 T_{\dd'}^2}4 +\frac{ t_{0\dd'}^2}{T_{\dd'}^2} + \ii\mu t_{0\dd'} } \text{erfi}\left(\frac1{\sqrt2} \left( \frac{\nu T_\dd-\mu T_{\dd'}}2 -\ii\frac{t_{0\dd}}{T_\dd} +\ii\frac{t_{0\dd'}}{T_{\dd'}} \right) \right).
\end{split}
\end{equation}
Inserting this above yields
\begin{equation}
     \integral{t'}{-\infty}\infty \ee{\ii\mu t'} \eta_{\dd'}(t')\integral{t}{-\infty}{t'}\ee{\ii\nu t}\eta_\dd(t) =\frac{T_\dd T_{\dd'}\pi}2  \ee{-\frac{\mu^2 T_{\dd'}^2+\nu^2 T_\dd^2}4}  \ee{\ii (\nu t_0\dd+\mu t_{0\dd'})}   \left( 1-\ii \,   
        \text{erfi}\left[  \frac{\nu T_\dd-\mu T_{\dd'}}{2\sqrt2} +\frac\ii{\sqrt2}\left(\frac{t_{0\dd'}}{T_{\dd'}} -\frac{t_{0\dd}}{T_{\dd}} \right) \right]
       \right).\label{eq:3rd Gaussian int}
\end{equation}

For stationary detectors we  use $\tau_\dd(t)=\sqrt{f(r_\dd)}t $ as the relation between proper time and coordinate time (see Eq.~\eqref{eq:static proper t}).
With this, and Eqs.~\eqref{eq:GF} and~\eqref{eq:2nd int Gauss}, the single detector noise term $L^\Psi_{\dd\dd'}$ in Eq.~\eqref{eq:L_integral_defn} takes the following form (denoting $r=r_\dd,r'=r_{\dd'}$,  $f_\dd=f(r_\dd)$ and $N_\dd=\sqrt{f_\dd}$):
\begin{equation}\label{eq:L_final_expression}
\begin{split}
    &L_{\dd\dd'}^\Psi= \lambda_\dd\lambda_{\dd'} f_\da f_\db \integral{t}{-\infty}\infty\integral{t'}{-\infty}\infty \eta_\dd(t) \eta_{\dd'}(t') \ee{-\ii\Omega_\dd N_\dd t+\ii\Omega_{\dd'} N_{\dd'}t' } \Wmanpsix{\coord x_\dd(t)}{\coord x_{\dd'}(t')}\Psi
    \\& 
    = \frac{\lambda_\dd\lambda_{\dd'} f_\da f_\db}{\left(4\pi\right)^2}\sum_{\ell=0}^{\infty}(2\ell+1)P_{\ell}(\cos\gamma) \integral{\omega}{-\infty}{\infty} \frac{G_{\ell\omega}^\Psi(r_\dd,r_{\dd'})}{\omega}  
    \left(\integral{t}{-\infty}\infty \eta_\dd(t)  \ee{-\ii (\Omega_\dd N_\dd+\omega)t}\right) \left(\integral{t'}{-\infty}\infty\eta_{\dd'}(t') \ee{ \ii (\Omega_{\dd'} N_{\dd'}+ \omega) t' }\right)
    \\& = 
    \frac{\lambda_\dd\lambda_{\dd'} N_\dd N_{\dd'} T_\dd T_{\dd'}}{16\pi }\sum_{\ell=0}^{\infty}(2\ell+1)P_{\ell}(\cos\gamma) \integral{\omega}{-\infty}{\infty} \frac{G_{\ell\omega}^\Psi(r_\dd,r_{\dd'})}{\omega}
    \ee{\ii(\Omega_{\dd'}N_{\dd'}+ \omega)t_{0\dd'}-\ii(\Omega_{\dd}N_{\dd}+ \omega)t_{0\dd}-\frac{1}{4} (\Omega_{\dd}N_{\dd}+ \omega)^2 T_{\dd}^2-\frac{1}{4} (\Omega_{\dd'}N_{\dd'}+ \omega)^2 T_{\dd'}^2} 
\end{split}
\end{equation}

Analogously, for $M^\Psi$ in Eq.~\eqref{eq:M_integral_defn} we obtain using Eq.~\eqref{eq:3rd Gaussian int},
\begin{equation}
\label{eq:M_final_expression}
\begin{split}
    &M^\Psi
	=\frac{-\lambda_\da\lambda_\db f_\da f_\db}{(4\pi)^2} \sum_{\ell=0}^{\infty}(2\ell+1)P_{\ell}(\cos\gamma) \integral{\omega}{-\infty}{\infty} \frac1\omega
    \left(  \integral{t}{-\infty}\infty \ee{\ii ( \Omega_\da f_\da- \omega)t}\eta_\da(t) \integral{t'}{-\infty}t \ee{\ii (\omega +\Omega_\db f_\db) t'} \eta_\db(t')  G_{\ell\omega}^\Psi(r_\da,r_{\db}) 
    \right.\\&\left.\qquad \qquad 
    +  \integral{t}{-\infty}\infty  \eta_\db(t)\ee{\ii( \Omega_\db f_\db -\omega) t}  \integral{t'}{-\infty}t \ee{\ii (\omega+ \Omega_\da f_\da) t'}\eta_\da(t')   G_{\ell\omega}^\Psi(r_\db,r_{\da})\right)
    \\
    &=\frac{-\lambda_\da\lambda_\db f_\da f_\db T_\da T_\db}{32\pi} \sum_{\ell=0}^{\infty}(2\ell+1)P_{\ell}(\cos\gamma) \integral{\omega}{-\infty}\infty \frac1\omega  
    \\&\quad
    \times\left( \ee{-\frac{\mu_\da^2 T_{\da}^2+\nu_\db^2 T_\db^2}4}  \ee{\ii (\nu_\db t_{0\db}+\mu_\da t_{0\da})}    \left( 1-\ii \,   \erfi\left[  \frac{\nu_\db T_\db-\mu_\da T_{\da}}{2\sqrt2} +\frac\ii{\sqrt2}\left(\frac{t_{0\da}}{T_{\da}} -\frac{t_{0\db}}{T_{\db}} \right) \right]  \right)  G_{\ell\omega}^\Psi(r_\da,r_{\db}) \right.
    \\&\left.\qquad \qquad +  
    \ee{-\frac{\mu_\db^2 T_{\db}^2+\nu_\da^2 T_\da^2}4}  \ee{\ii (\nu_\da t_{0\da}+\mu_\db t_{0\db})}   \left( 1-\ii \,   
    \erfi\left[  \frac{\nu_\da T_\da-\mu_\db T_{\db}}{2\sqrt2} +\frac\ii{\sqrt2}\left(\frac{t_{0\db}}{T_{\db}} -\frac{t_{0\da}}{T_{\da}} \right) \right]
       \right) G_{\ell\omega}^\Psi(r_\db,r_{\da}) \right),
\end{split}
\end{equation}
with
$
\nu_\dd= \omega+\Omega_\dd N_\dd,\, \mu_\dd=\Omega_\dd N_\dd-\omega
$. For practical evaluation purposes, one can manipulate the integrals in Eqs.~\eqref{eq:M_final_expression} and~\eqref{eq:L_final_expression} further so that the integration over $\omega$ only runs over $0<\omega<\infty$.
\end{widetext}

\section{Numerical techniques for the Wightman function}\label{app:num techniques}

\begin{figure*}
\centering
\subfloat[][Integrand in Eq.~\eqref{eq: Worked L term integral}]{
\includegraphics[width = 0.95\columnwidth]{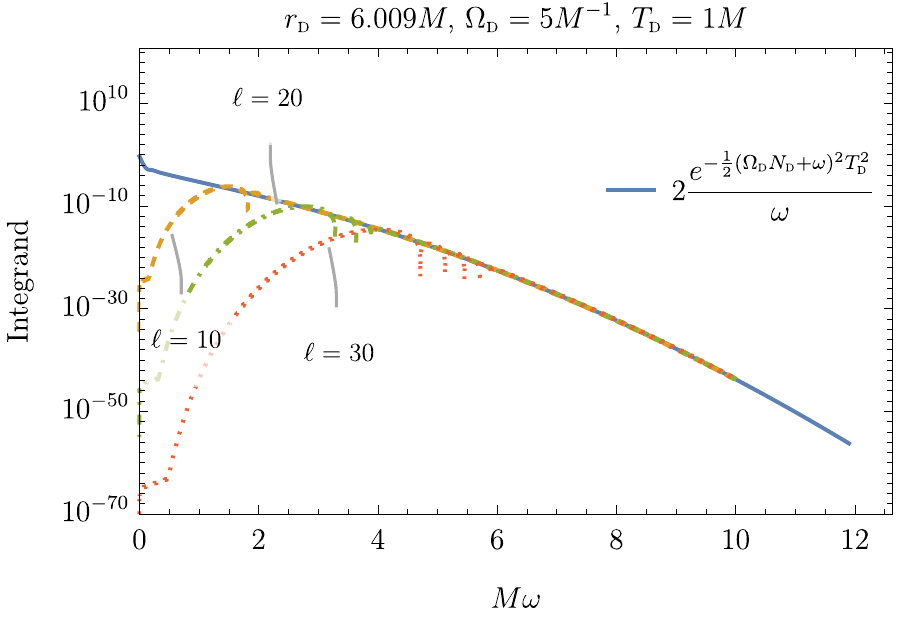}
\label{plt:lTermIntegrand}}
\subfloat[][Relative difference integral in Eq.~\eqref{eq: Worked L term integral}]{
\includegraphics[width = 0.95\columnwidth]{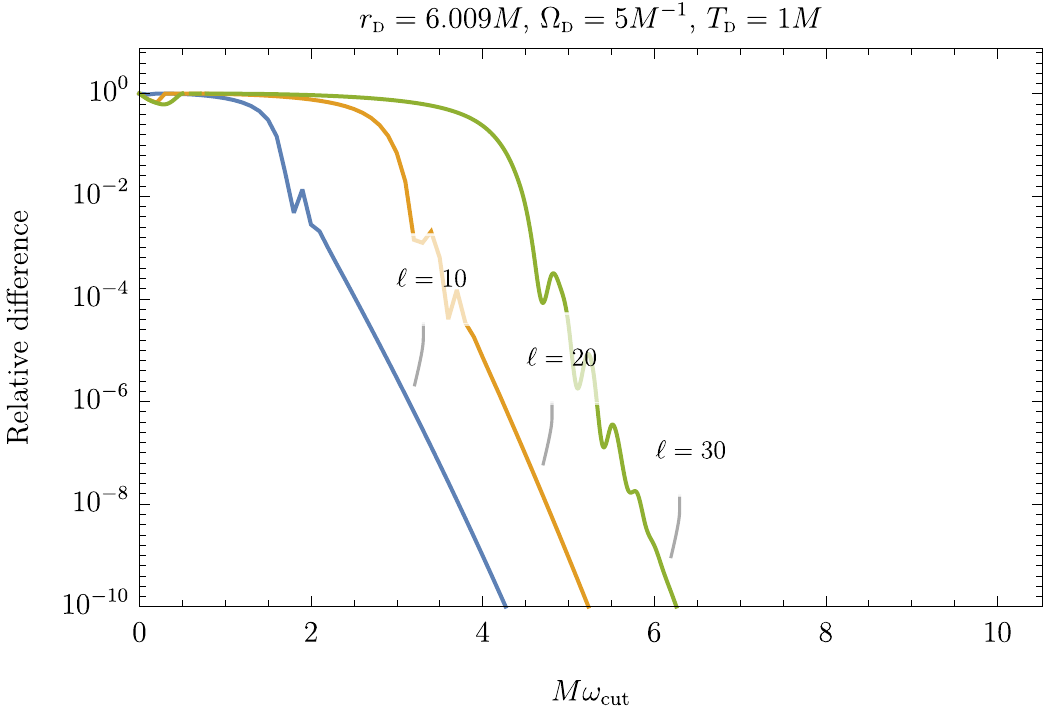}
\label{plt:lTermIntegralConvergence}}
\qquad
\subfloat[][Summand in Eq.~\eqref{eq: Worked L term integral}]{
\includegraphics[width = 0.95\columnwidth]{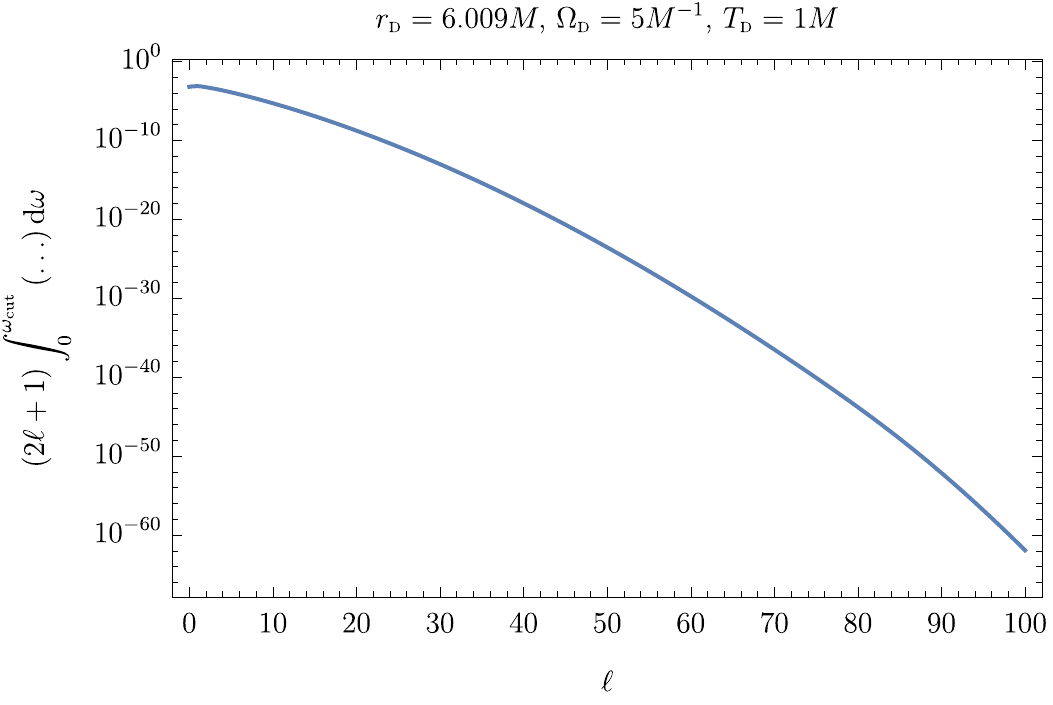}
\label{plt:LSummandConvergence}}
\subfloat[][Relative difference $\ell$-sum in Eq.~\eqref{eq: Worked L term integral}]{
\includegraphics[width = 0.95\columnwidth]{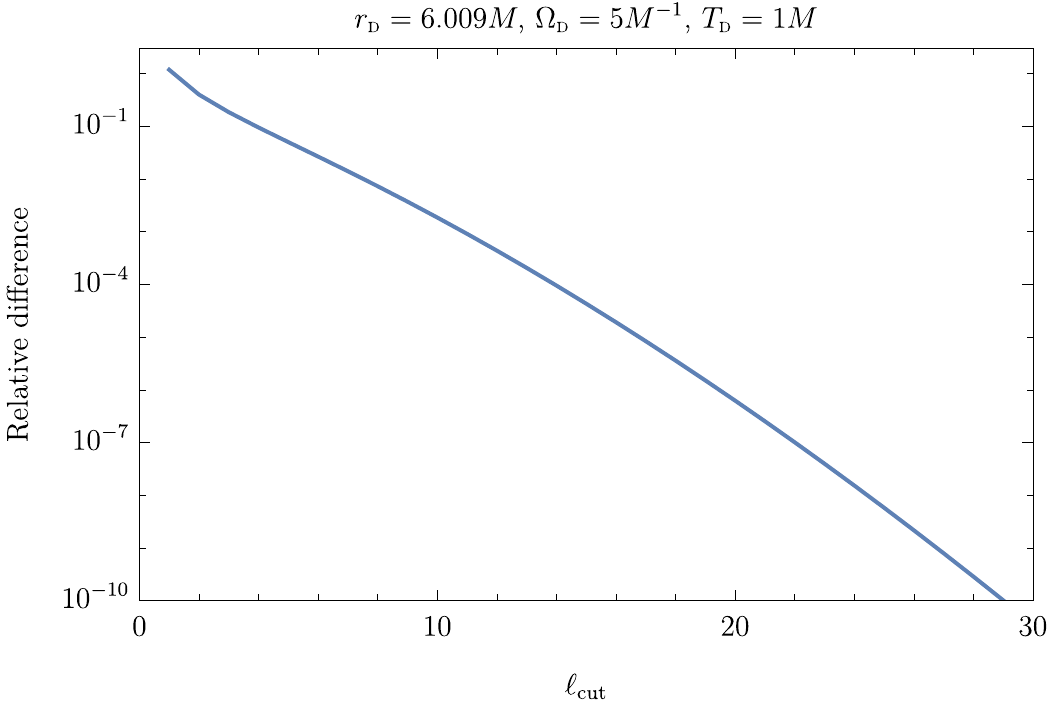}
\label{plt:LSumConvergence}}
\caption{(a) Integrand in Eq.~\eqref{eq: Worked L term integral} as a function of $M\omega$ and $\ell$ for the Boulware state. The value of $M\omega$ where the asymptotic regime begins grows with $\ell$. More precisely, the transition to the asymptotic regime begins around $\omega r \sim \sqrt{\ell\left(\ell+1\right)}$ (for $\ell>0$ and radii not too close to the horizon). Notice that thanks to the $\theta(\omega)$ in Eq.~\eqref{eq:FGF modes B}, the slowest decaying term in Eq.~\eqref{eq: Worked L term integral} for the Boulware state is proportional to $\sim 2\frac{\ee{-\frac{1}{2} (\Omega_\dd N_\dd + \omega)^2 T_D^2}}{\omega}$.
\\
(b) Relative difference  between the integral in Eq.~\eqref{eq: Worked L term integral} for the Boulware state when integrated up to $\omega_\mathrm{cut}$ and up to $\omega_\mathrm{cut}+\delta$, with $\delta = 1/10$. The reason to limit this plot to $10^{-10}$ is that those integrals where evaluated to ten significant digits; hence, a relative difference of $10^{-10}$ means that the integral converged completely to all significant digits. Comparing with Fig.~\ref{plt:lTermIntegrand}, we see that exponentially fast convergence happens after $M\omega_\mathrm{cut}$ becomes slightly larger than the $M\omega$ where the asymptotic regime begins.
\\
(c) Summand in Eq.~\eqref{eq: Worked L term integral} as a function of $\ell$ for the Boulware state. One can notice from this figure that the summand decays superexponentially with increasing $\ell$. In fact, for $\ell > 30$, it has already converged to all 10 significant digits that we got when evaluating the integrals.
\\
(d) Relative difference between the $\ell$-sum in Eq.~\eqref{eq: Worked L term integral} for the Boulware state when summed up to $\l_\mathrm{cut}$ and $\l_\mathrm{cut}-1$. The reason to limit this plot to $10^{-10}$ is that the integrals in the sums were to 10 digits of precision. Hence, a relative difference that is smaller than $10^{-10}$ should represent numerical noise and not actual significant digits.
}
\end{figure*}

In order to evaluate the Wightman function in the situations presented in this work, integrals over $\Rinup(r)$ with respect to $r$ are needed. 
However, there is no known closed-form expression for $\Rinup(r)$. Hence, one has to use numerical techniques. Specifically, we use numerical methods to evaluate $\Rin$, $\Rup$ and the set of coefficients $I_{\ell\omega}, \rho^{in/up}_{\ell\omega}$ in the region $M \omega\in\left[0,10\right]$ in steps of $10^{-3}$, $r^*/M\in\left[-4,13\right]$ in steps of $1/5$ and $\ell$ for all (integer) values $0\leq\ell\leq100$. Different techniques are involved in the evaluation.

For $\Rup$ we used the NDSolve-based numerical method implemented in the Regge-Wheeler Mathematica~\cite{Mathematica} package of the Black Hole Perturbation Toolkit~\cite{BHPToolkit}.

For $\Rin$ we define two auxiliary quantities $\bar{r} = r/2M$ and $\bar{\omega} = 2M\omega$, then use the so-called Jaffé series~\cite{Leaver1986} to write the $\Rin$ solution to~\eqref{eq:regge_wheeler} for $\bar{r}\in\left[1,\infty\right)$ as
\begin{equation}\label{eq:Jaffe}
    \Rin\left(\bar{r}\right) = \bar{r}^{2\ii\bar{\omega}}\left(\bar{r}-1\right)^{-\ii\bar{\omega}}\ee{\ii\bar{\omega}\bar{r}}\sum_{n=0}^{\infty}a_{n}\left(\frac{\bar{r}-1}{\bar{r}}\right)^n,
\end{equation}
where
\begin{align}
    &a_n\alpha_{n-1} + a_{n-1}\beta_{n-1}+a_{n-2}\gamma_{n-1}=0,\\
    &\alpha_{n}=(n+1)(n+1-2\ii\bar{\omega}),\nonumber\\
    &\beta_{n} = -1-2n(n+1)-\ell(\ell+1)+4\bar{\omega}\left(\ii+2n\ii+2\bar{\omega}\right),\nonumber\\
    &\gamma_{n} = (n-2\ii\bar{\omega})^2,\nonumber
\end{align}
with $a_0 = \ee{-2\ii\bar{\omega}}$ and $a_n = 0\,\forall\, n < 0$.
This solution is implemented in Mathematica. However some care has to be taken: in practice we have to impose a cutoff on the number of terms $n_{max}$ to include in the series. To set an adequate value for $n_{max}$ we used the following quantity:
\begin{align}\label{eq:Jaffe series error}
    \varepsilon\left(\ell,\bar{\omega},\bar{r},n_{max}\right) = \frac{a_{n_{max}+1}\left(\frac{\bar{r}-1}{\bar{r}}\right)^{n_{max}+1}}{\sum_{n=0}^{n_{max}}a_{n}\left(\frac{\bar{r}-1}{\bar{r}}\right)^n}.
\end{align}
Since we want to have at least $16$ digits of precision in the $\Rin$ modes, we want to find an $n_{max}$ such that
\begin{equation}
    \begin{aligned}\label{eq:Jaffe series condition}
        &\Re\left[\varepsilon\left(\ell,\bar{\omega},\bar{r},n_{max}\right)\right] <10^{-16}\text{ and}\\
        &\Im\left[\varepsilon\left(\ell,\bar{\omega},\bar{r},n_{max}\right)\right] <10^{-16}
    \end{aligned}
\end{equation}
 for all values of $\bar{r}$,$\ell$ and $\bar{\omega}$ in the region where the solutions are evaluated. For that, we use the fact that the convergence of the series becomes slower with increasing $\bar{r}$ to conclude that the largest $\varepsilon$ happens at the highest $\bar{r}$ we intended to evaluate, which is around $\bar{r}_{max}\approx5$. Then, fixing $\bar{r} = \bar{r}_{max}$ we test a couple of $\bar\omega$ and $\ell$ values in the region of interest and conclude that $n_{max} = 5000$ is enough to satisfy Eq.~\eqref{eq:Jaffe series condition} in all regions of interest.

For the $I_{\ell\omega}, \rho^{in/up}_{\ell\omega}$ coefficients we use data from Ref.~\cite{BUSS2018168}, which is evaluated using a Mathematica implementation of the Mano-Suzuki-Takasugi method. An extensive review of that method can be found in Ref.~\cite{Sasaki2003}.

\subsection{$L_{\dd\dd}^\Psi$-terms}
By setting $\dd = \dd'$ in Eq.~\eqref{eq:L_final_expression} and manipulating the integration range from $-\infty<\omega<\infty$ to $0<\omega<\infty$, we obtain the following expression for the $L_{\dd\dd}^\Psi$-terms:
\begin{widetext}
\begin{align}\label{eq: Worked L term integral}
    L_{\dd\dd}^\Psi\left(r_\dd\right) = \frac{\left( N_\dd \lambda_\dd T_\dd \right)^2}{16\pi}\sum_{\ell=0}^{\infty}(2\ell+1)
\integral{\omega}{0}{\infty} \frac{1}{\omega}\left(G_{\ell\omega}^\Psi(r_\dd,r_{\dd})\ee{-\frac{1}{2} (\Omega_\dd N_\dd + \omega)^2 T_D^2}-G_{\ell,-\omega}^\Psi(r_\dd,r_{\dd})\ee{-\frac{1}{2} (\Omega_\dd N_\dd - \omega)^2 T_D^2}\right).
\end{align}
To verify the convergence of such an integral, one should check the large-$|\omega|$ behavior of the $G_{\ell\omega}^{\Psi}(r_\dd,r_\dd)$, which varies among the quantum states~\eqref{eq:FGF modes B}-\eqref{eq:FGF modes H}:
\begin{align}\label{eq:G large-w}
    &G_{\ell\omega}^{B}(r_\dd,r_\dd) \sim \theta(\omega)\left(|\RupC(r_\dd)|^2 + |\RinC(r_\dd)|^2\right),\\
    &G_{\ell\omega}^{U}(r_\dd,r_\dd) \sim |\RupC(r_\dd)|^2\left(\theta(\omega)-\theta(-\omega)\ee{2\pi\omega\kappa}\right) + \theta(\omega)|\RinC(r_\dd)|^2,\nonumber\\
    &G_{\ell\omega}^{H}(r_\dd,r_\dd) \sim |\RupC(r_\dd)|^2\left(\theta(\omega)-\theta(-\omega)\ee{2\pi\omega\kappa}\right) + |\RinC(r_\dd)|^2\left(\theta(\omega)\ee{-2\pi\omega\kappa}-\theta(-\omega)\right).\nonumber
\end{align}
\end{widetext}
The leading asymptotic behavior of $|\RinupC|^2$ as $|\omega|\to\infty$ is $|\RinupC|^2 \sim 1$~\cite{Levi2016}\footnote{Such asymptotics are nonuniform and valid only when ${\omega \gg V_{\ell}(r)}$, as illustrated in Fig.~\ref{plt:lTermIntegrand}}. 
Hence, to leading order for large $|\omega|$, $G_{\ell\omega}^{\Psi}(r_\dd,r_\dd) \sim 1$. 
By substituting such a result into the integrand in Eq.~\eqref{eq: Worked L term integral}, one concludes that the slowest-decaying term of the integrand falls off as $\sim 2\frac{\ee{-\frac{1}{2} (\Omega_\dd N_\dd - \omega)^2 T_D^2}}{\omega}$. Therefore, the integral in Eq.~\eqref{eq: Worked L term integral} is fast-converging. Given that, we are able to numerically evaluate it accurately enough with a frequency cutoff $M\omega_{cut} = 10$. The asymptotic regime of the integrand and the convergence of the integral are illustrated in Figs.~\ref{plt:lTermIntegrand} and~\ref{plt:lTermIntegralConvergence}.

In order to verify the convergence of the $\ell$-sum, we begin by evaluating the integrals in Eq.~\eqref{eq: Worked L term integral}, multiplying by $2\ell + 1$ and plotting the result against $\ell$. The outcome is shown in Fig.~\ref{plt:LSummandConvergence}, where we can see that such a quantity decays superexponentially. Given this numerical evidence, the $\ell$-sum in that equation is expected to converge with a good accuracy even with a cutoff lower than the $\ell_\mathrm{cut}=100$ we use. In fact, as presented in Fig.~\ref{plt:LSumConvergence}, $\ell_{\mathrm{cut}} = 30$ is enough to converge to all ten significant digits used when evaluating the Boulware integrals.

\begin{figure*}
\centering
\subfloat[][Boulware state integrand in Eq.~\eqref{eq:M_final_expression}]{
\includegraphics[width = .95\columnwidth]{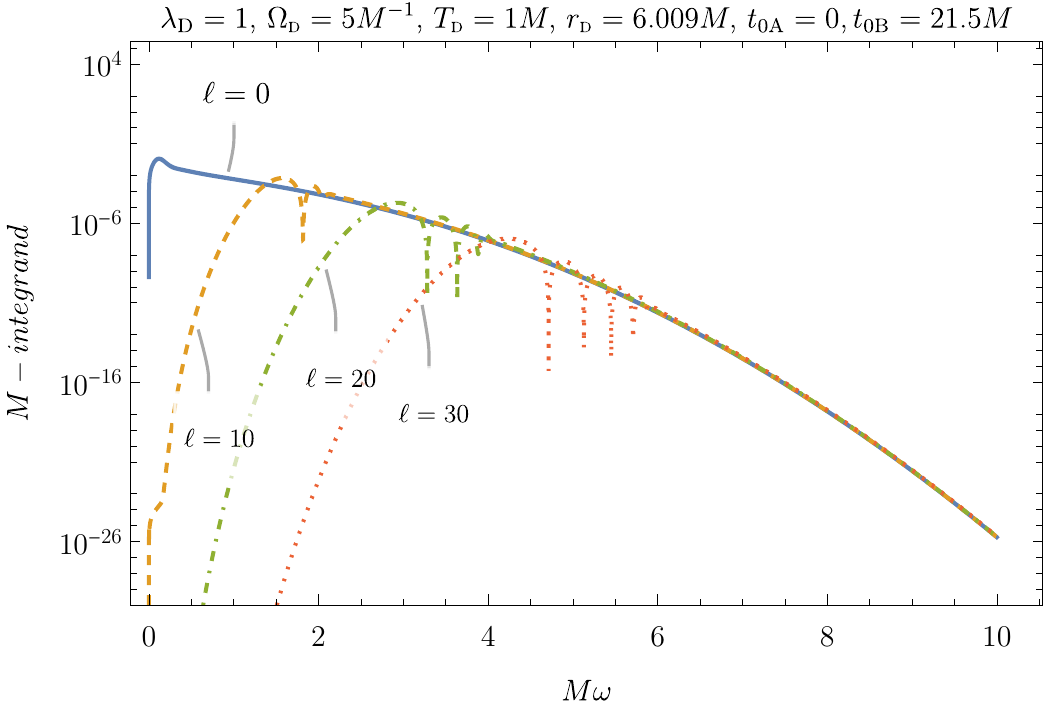}
\label{fig: M-integrand}}
\subfloat[][Relative difference Boulware  integral in Eq.~\eqref{eq:M_final_expression}]{
\includegraphics[width =.95 \columnwidth]{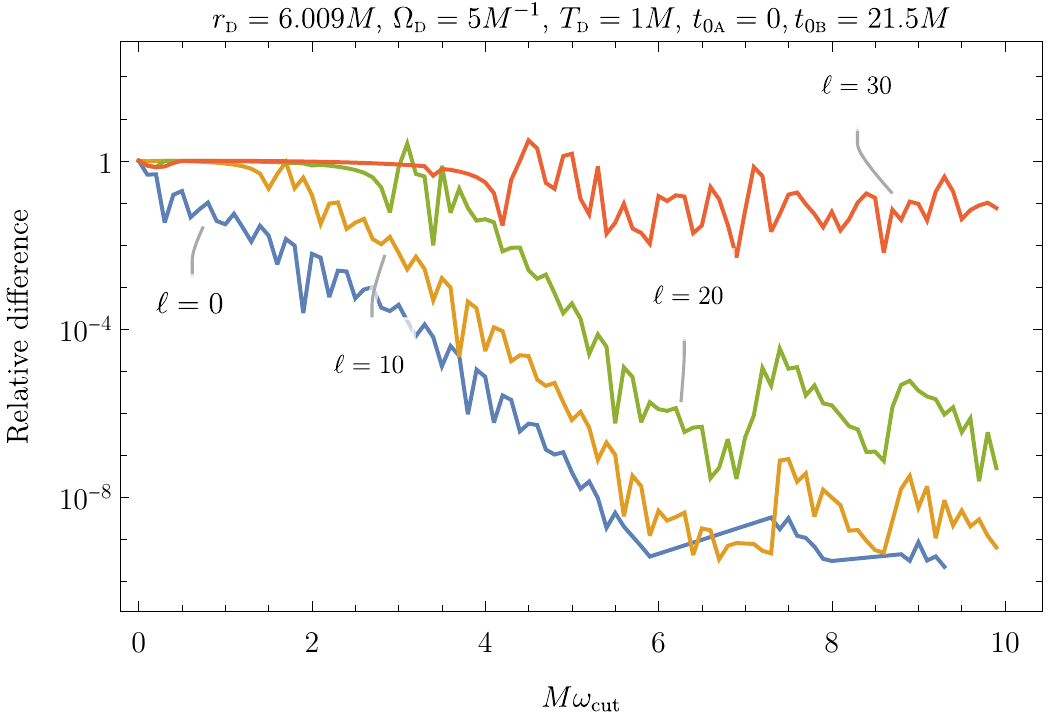}
\label{fig: M-integral}}
\qquad
\subfloat[][Boulware state integral in Eq.~\eqref{eq:M_final_expression}]{
\includegraphics[width = .95\columnwidth]{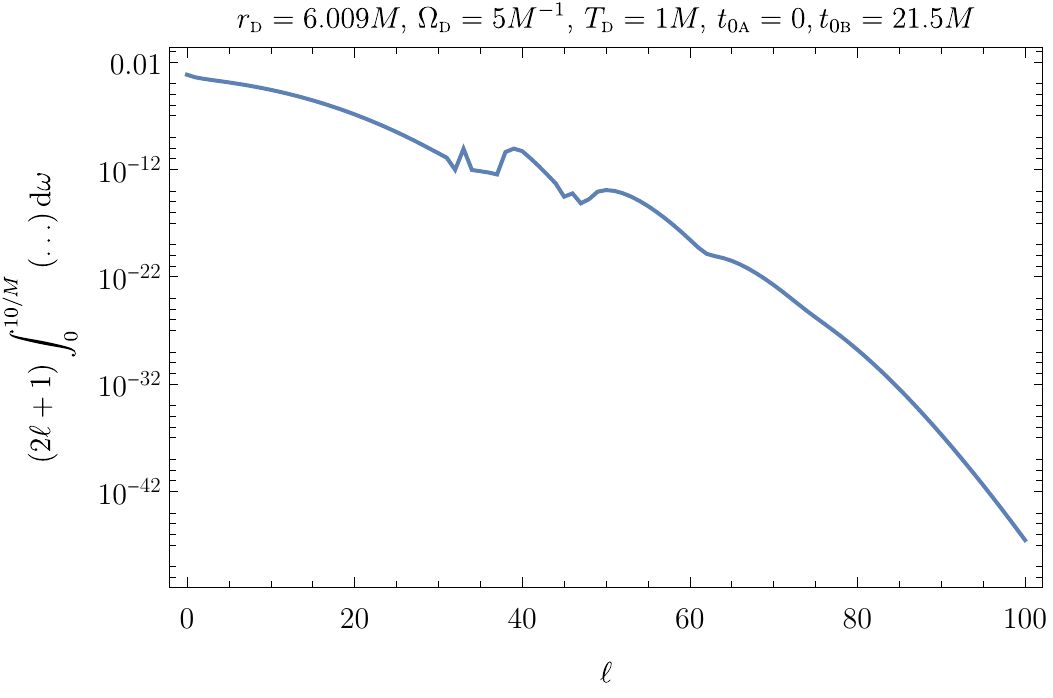}
\label{fig: M-integrals up to l 100}}
\subfloat[][Relative difference Boulware integral from~\eqref{eq:M_final_expression}]{
\includegraphics[width = .95\columnwidth]{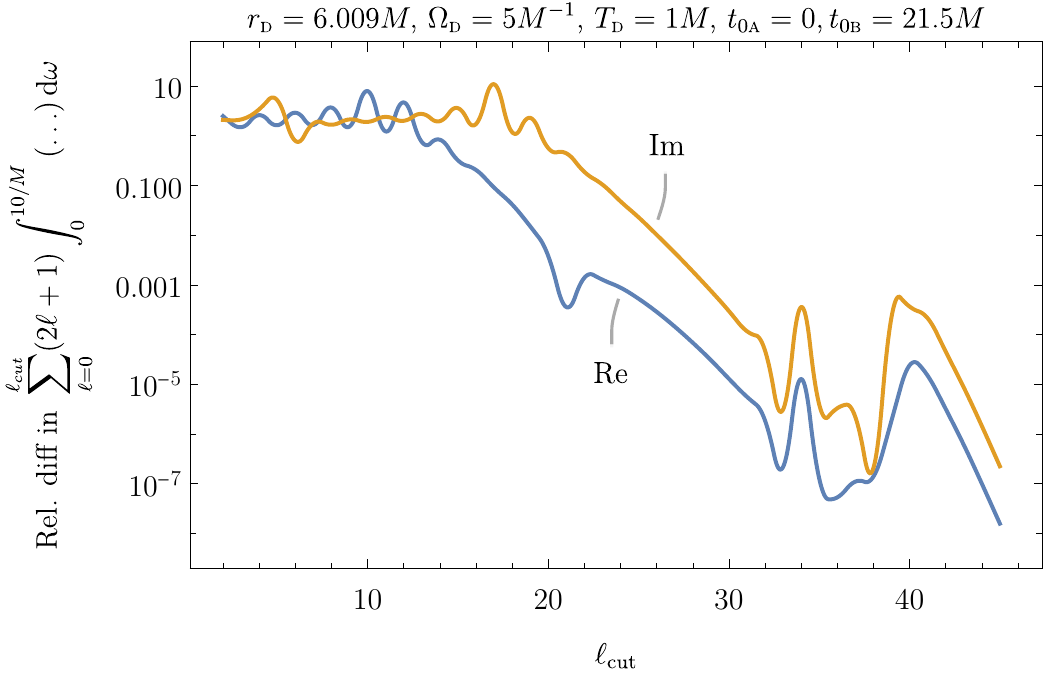}
\label{fig: M-integrals relative difference}}
\caption{(a)  Boulware state integrand in Eq.~\eqref{eq:M_final_expression} after rewriting the integral so that it ranges from $\omega = 0$ to $M\omega = 10$, as a function of $M\omega$, for several different $\ell = 0,10,20,30$.
\\
(b) Relative difference in the Boulware state integral in Eq.~\eqref{eq:M_final_expression} after rewriting the integral so that it ranges from $\omega = 0$ to $\omega = \omega_{cut}$ for several different $\ell = 0,10,20,30$. To evaluate the relative difference, we integrate up to $\omega_{cut}$ and then to $\omega_{cut}+\delta$, with $\delta=1/10$.
\\
(c) Boulware state integral in Eq.~\eqref{eq:M_final_expression}, after rewriting the integral so that it ranges from $\omega = 0$ to $\omega = \omega_{cut}$ for several different $\ell = 0,10,20,30$, here integrated up to $M\omega_{cut}=10$.
\\
(d) Relative difference in the Boulware state integral from~\eqref{eq:M_final_expression}, after rewriting the integral so that it ranges from $\omega = 0$ to $\omega = \omega_{cut}$ for several different $\ell = 0,10,20,30$. To evaluate the relative difference, we sum up to $\ell_{cut}$ and then to $\ell_{cut}+1$, ranging form $\ell_{cut} = 1$ up to $\ell_{cut} = 100$. For $\ell_{cut} > 45$ all significant digits exactly cancel out when evaluating the relative difference.
}
\end{figure*}

\subsection{$M^\Psi$-terms}
Since we already know the large-$|\omega|$ asymptotics of $G_{\ell\omega}^{\Psi}(r_\dd,r_\dd)$, to analyze the behavior of the integrand in Eq.~\eqref{eq:M_final_expression} in that regime we have to study the factors that multiply each of its terms. 
These factors are very similar and it suffices to analyze only one of them:
\begin{widetext}
\begin{equation}
    \ee{-\frac{\mu_\da^2 T_{\da}^2+\nu_\db^2 T_\db^2}4}\ee{\ii (\nu_\db t_{0\db}+\mu_\da t_{0\da})}
    \erfc\left[  \frac{\ii (\nu_\db T_\db-\mu_\da T_{\da})}{2\sqrt2} -\frac{1}{\sqrt2}\left(\frac{t_{0\da}}{T_{\da}} -\frac{t_{0\db}}{T_{\db}} \right) \right],
\end{equation}
where $\text{erfc}(z)=1-\text{erf}(z)$. Expanding the exponential functions, we obtain
\begin{equation}
    \alpha_{\da \db}
    \ee{-\frac{1}{4}(T_{\da}^2+T_\db^2)\omega^2}
    \ee{-\frac{1}{2}\left(N_{\db}T_{\db}^2\Omega_{\db}-N_{\da}T_{\da}^2\Omega_{\da}\right)\omega}
    \ee{\ii\omega(t_{0\db}- t_{0\da})}
    \erfc\left[  \frac{\ii (\nu_\db T_\db-\mu_\da T_{\da})}{2\sqrt2} -\frac{1}{\sqrt2}\left(\frac{t_{0\da}}{T_{\da}} -\frac{t_{0\db}}{T_{\db}} \right) \right],
\end{equation}
where
\begin{equation}
    \alpha_{\da \db}=\ee{-\frac{1}{4}\left(N_{\db}^2 T_{\db}^2 \Omega_{\db}^2 + N_{\da}^2 T_{\da}^2 \Omega_{\da}^2\right)}
    \ee{\ii (\Omega_{\db}N_{\db} t_{0\db}+\Omega_{\da}N_{\da} t_{0\da})}.
\end{equation}
Now we use an asymptotic expression for the $\erfc(z)$,
\begin{equation}
    \erfc(z)\sim \frac{e^{-z^2}}{z\sqrt{\pi}},
    \quad
    z\to\infty,\quad |\text{ph}(z)|<3\pi/4,
\end{equation}
to conclude that as $\omega\to\infty$,
\begin{align}\label{eq: erfc asymptotics}
    &\erfc\left[  \frac{\ii (\nu_\db T_\db-\mu_\da T_{\da})}{2\sqrt2} -\frac{1}{\sqrt2}\left(\frac{t_{0\da}}{T_{\da}} -\frac{t_{0\db}}{T_{\db}} \right) \right]
    \sim
    \beta_{\da \db}
    \frac{\ee{\frac{1}{8}(T_{\da}+T_{\db})^2\omega^2}
    \ee{\frac{1}{4}(T_\da + T_\db)(N_\db T_\db \Omega_\db - N_\da T_\da \Omega_\da)\omega}
    \ee{-\ii\omega\frac{1}{2}(T_\da + T_\db)\left(\frac{t_{0\db}}{T_\db}-\frac{t_{0\da}}{T_\da}\right)}}{\frac{\ii ((N_\db \Omega_\db + \omega) T_\db - (N_\da \Omega_\da - \omega) T_{\da})}{2\sqrt2} -\frac{1}{\sqrt2}\left(\frac{t_{0\da}}{T_{\da}} -\frac{t_{0\db}}{T_{\db}} \right)},
\end{align}
where
\begin{equation}
    \beta_{\da \db} = 
    (\pi\alpha_{\da \db})^{-1/2}
    \ee{-\frac{1}{2}(\frac{t_{0\db}}{ T_\db} - \frac{t_{0\da}}{ T_\da})^2}
    \ee{\frac{\ii}{2} (N_{\da} T_{\da} \Omega_{\da} \frac{t_{0\db}}{T_\db} + N_{\db} T_{\db} \Omega_{\db} \frac{t_{0\da}}{T_\da})}
    \ee{-\frac{1}{4}(N_{\da} T_{\da} \Omega_{\da} N_{\db} T_{\db} \Omega_{\db})}.
\end{equation}
Putting everything together we conclude that
\begin{equation}\begin{split}\label{eq: m-term error function asymptotics}
    &\ee{-\frac{\mu_\da^2 T_{\da}^2+\nu_\db^2 T_\db^2}4}\ee{\ii (\nu_\db t_{0\db}+\mu_\da t_{0\da})}
    \erfc\left[  \frac{\ii (\nu_\db T_\db-\mu_\da T_{\da})}{2\sqrt2} -\frac{1}{\sqrt2}\left(\frac{t_{0\da}}{T_{\da}} -\frac{t_{0\db}}{T_{\db}} \right) \right]\sim
    \\
    &\quad\quad\quad
    \alpha_{\da \db}
    \beta_{\da \db}
    \frac{
    \ee{-\frac{1}{8}(T_{\da}-T_\db)\omega^2}
    \ee{\frac{1}{4}(T_\da - T_\db)(N_\db T_\db \Omega_\db + N_\da T_\da \Omega_\da)\omega}
    \ee{\frac{\ii}{2}\omega(T_{\db}- T_{\da})\left(\frac{t_{0\da}}{T_\da}+\frac{t_{0\db}}{T_\db}\right)}
    }
    {\frac{\ii  ((N_\db \Omega_\db + \omega) T_\db - (N_\da \Omega_\da - \omega) T_{\da})}{2\sqrt2} -\frac{1}{\sqrt2}\left(\frac{t_{0\da}}{T_{\da}} -\frac{t_{0\db}}{T_{\db}} \right)},
    \quad \omega\to\infty,
\end{split}\end{equation}
which leads to the conclusion that, whenever $T_{\da}\neq T_\db$, the whole term decays superexponentially as $\ee{-\frac{1}{8}(T_{\db}-T_\da)^2\omega^2}$. On the other hand, if $T_\da = T_\db \equiv T$, which is the case we study in this work,~\eqref{eq: m-term error function asymptotics} simplifies to 
\begin{equation}
\label{eq:int TA=TB}
\begin{split}
     &\ee{-\frac{(\mu_\da^2+\nu_\db^2)T^2}{4}}\ee{\ii (\nu_\db t_{0\db}+\mu_\da t_{0\da})}
    \erfc\left[  \frac{\ii (\nu_\db-\mu_\da)T}{2\sqrt2} +\frac{1}{T\sqrt2}\left(t_{0\db} -t_{0\da} \right) \right]
    \sim
    \frac{\alpha_{\da \db}\beta_{\da \db}}{\frac{\ii  (N_\db \Omega_\db - N_\da \Omega_\da + 2\omega)T}{2\sqrt2} -\frac{(t_{0\da} -t_{0\db})}{T\sqrt2}},
\quad \omega\to\infty.
\end{split}\end{equation}
\end{widetext}

Then both, the real and imaginary parts of this quantity decay linearly with $\omega$, since the absolute value of the denominator grows linearly with the frequency. Given that, by considering the extra $\omega^{-1}$ in the integrand in Eq.~\eqref{eq:M_final_expression}, we conclude that it decays as $\omega^{-2}$. 
Hence, the integrals defining the $M^\Psi$-terms are convergent.

We remark that such leading asymptotic behavior in Eq.~\eqref{eq:int TA=TB} is, in general, not achieved by $M\omega = 10$, but one can still perform the numerical integration to a good accuracy because, before this leading asymptotic regime, there is an intermediary regime where the integrand is exponentially decaying, as can be seen in Fig.~\ref{fig: M-integrand}. 
We can further confirm that convergence by looking at the integral in Eq.~\eqref{eq:M_final_expression} as a function of $\omega_{cut}$: as can be seen in Fig.~\ref{fig: M-integral}, up to $\ell = 20$ we have convergence up to four significant digits, while for larger values of $\ell$, the convergence becomes worse, resulting in no significant digits at all. Yet, since the resulting integral decays superexponentially with $\ell$, as presented in Fig.~\ref{fig: M-integrals up to l 100}, one can still obtain up to four significant digits, as can be seen in Fig.~\ref{fig: M-integrals relative difference}.

\bibliography{entanglement_harvesting_schwarzschild}

\begin{thebibliography}{51}%
\makeatletter
\providecommand \@ifxundefined [1]{%
 \@ifx{#1\undefined}
}%
\providecommand \@ifnum [1]{%
 \ifnum #1\expandafter \@firstoftwo
 \else \expandafter \@secondoftwo
 \fi
}%
\providecommand \@ifx [1]{%
 \ifx #1\expandafter \@firstoftwo
 \else \expandafter \@secondoftwo
 \fi
}%
\providecommand \natexlab [1]{#1}%
\providecommand \enquote  [1]{``#1''}%
\providecommand \bibnamefont  [1]{#1}%
\providecommand \bibfnamefont [1]{#1}%
\providecommand \citenamefont [1]{#1}%
\providecommand \href@noop [0]{\@secondoftwo}%
\providecommand \href [0]{\begingroup \@sanitize@url \@href}%
\providecommand \@href[1]{\@@startlink{#1}\@@href}%
\providecommand \@@href[1]{\endgroup#1\@@endlink}%
\providecommand \@sanitize@url [0]{\catcode `\\12\catcode `\$12\catcode
  `\&12\catcode `\#12\catcode `\^12\catcode `\_12\catcode `\%12\relax}%
\providecommand \@@startlink[1]{}%
\providecommand \@@endlink[0]{}%
\providecommand \url  [0]{\begingroup\@sanitize@url \@url }%
\providecommand \@url [1]{\endgroup\@href {#1}{\urlprefix }}%
\providecommand \urlprefix  [0]{URL }%
\providecommand \Eprint [0]{\href }%
\providecommand \doibase [0]{http://dx.doi.org/}%
\providecommand \selectlanguage [0]{\@gobble}%
\providecommand \bibinfo  [0]{\@secondoftwo}%
\providecommand \bibfield  [0]{\@secondoftwo}%
\providecommand \translation [1]{[#1]}%
\providecommand \BibitemOpen [0]{}%
\providecommand \bibitemStop [0]{}%
\providecommand \bibitemNoStop [0]{.\EOS\space}%
\providecommand \EOS [0]{\spacefactor3000\relax}%
\providecommand \BibitemShut  [1]{\csname bibitem#1\endcsname}%
\let\auto@bib@innerbib\@empty
\bibitem [{\citenamefont {Summers}\ and\ \citenamefont
  {Werner}(1985)}]{vacuumEntanglement}%
  \BibitemOpen
  \bibfield  {author} {\bibinfo {author} {\bibfnamefont {S.~J.}\ \bibnamefont
  {Summers}}\ and\ \bibinfo {author} {\bibfnamefont {R.}~\bibnamefont
  {Werner}},\ }\bibfield  {title} {\enquote {\bibinfo {title} {The vacuum
  violates bell's inequalities},}\ }\href {\doibase
  https://doi.org/10.1016/0375-9601(85)90093-3} {\bibfield  {journal} {\bibinfo
   {journal} {Phys. lett., A}\ }\textbf {\bibinfo {volume} {110}},\ \bibinfo
  {pages} {257--259} (\bibinfo {year} {1985})}\BibitemShut {NoStop}%
\bibitem [{\citenamefont {Summers}\ and\ \citenamefont
  {Werner}(1987)}]{vacuumBell}%
  \BibitemOpen
  \bibfield  {author} {\bibinfo {author} {\bibfnamefont {S.~J.}\ \bibnamefont
  {Summers}}\ and\ \bibinfo {author} {\bibfnamefont {R.}~\bibnamefont
  {Werner}},\ }\bibfield  {title} {\enquote {\bibinfo {title} {Bell’s
  inequalities and quantum field theory. i. general setting},}\ }\href
  {\doibase 10.1063/1.527733} {\bibfield  {journal} {\bibinfo  {journal}
  {Journal of Mathematical Physics}\ }\textbf {\bibinfo {volume} {28}},\
  \bibinfo {pages} {2440--2447} (\bibinfo {year} {1987})}\BibitemShut {NoStop}%
\bibitem [{\citenamefont {Saravani}\ \emph {et~al.}(2016)\citenamefont
  {Saravani}, \citenamefont {Aslanbeigi},\ and\ \citenamefont
  {Kempf}}]{saravani2016spacetime}%
  \BibitemOpen
  \bibfield  {author} {\bibinfo {author} {\bibfnamefont {M.}~\bibnamefont
  {Saravani}}, \bibinfo {author} {\bibfnamefont {S.}~\bibnamefont
  {Aslanbeigi}}, \ and\ \bibinfo {author} {\bibfnamefont {A.}~\bibnamefont
  {Kempf}},\ }\bibfield  {title} {\enquote {\bibinfo {title} {Spacetime
  curvature in terms of scalar field propagators},}\ }\href@noop {} {\bibfield
  {journal} {\bibinfo  {journal} {Physical Review D}\ }\textbf {\bibinfo
  {volume} {93}},\ \bibinfo {pages} {045026} (\bibinfo {year}
  {2016})}\BibitemShut {NoStop}%
\bibitem [{\citenamefont {Kempf}(2021)}]{kempf2021replacing}%
  \BibitemOpen
  \bibfield  {author} {\bibinfo {author} {\bibfnamefont {A.}~\bibnamefont
  {Kempf}},\ }\bibfield  {title} {\enquote {\bibinfo {title} {Replacing the
  notion of spacetime distance by the notion of correlation},}\ }\href@noop {}
  {\bibfield  {journal} {\bibinfo  {journal} {Frontiers in Physics}\ }\textbf
  {\bibinfo {volume} {9}},\ \bibinfo {pages} {655857} (\bibinfo {year}
  {2021})}\BibitemShut {NoStop}%
\bibitem [{\citenamefont {Perche}\ and\ \citenamefont
  {Mart\'{\i}n-Mart\'{\i}nez}(2022)}]{TalesEduSpacetimefromCorrelators}%
  \BibitemOpen
  \bibfield  {author} {\bibinfo {author} {\bibfnamefont {T.~R.}\ \bibnamefont
  {Perche}}\ and\ \bibinfo {author} {\bibfnamefont {E.}~\bibnamefont
  {Mart\'{\i}n-Mart\'{\i}nez}},\ }\bibfield  {title} {\enquote {\bibinfo
  {title} {Geometry of spacetime from quantum measurements},}\ }\href {\doibase
  10.1103/PhysRevD.105.066011} {\bibfield  {journal} {\bibinfo  {journal}
  {Phys. Rev. D}\ }\textbf {\bibinfo {volume} {105}},\ \bibinfo {pages}
  {066011} (\bibinfo {year} {2022})}\BibitemShut {NoStop}%
\bibitem [{\citenamefont {Bekenstein}(1973)}]{BenensteinEntropy1973}%
  \BibitemOpen
  \bibfield  {author} {\bibinfo {author} {\bibfnamefont {J.~D.}\ \bibnamefont
  {Bekenstein}},\ }\bibfield  {title} {\enquote {\bibinfo {title} {Black holes
  and entropy},}\ }\href {\doibase 10.1103/PhysRevD.7.2333} {\bibfield
  {journal} {\bibinfo  {journal} {Phys. Rev. D}\ }\textbf {\bibinfo {volume}
  {7}},\ \bibinfo {pages} {2333--2346} (\bibinfo {year} {1973})}\BibitemShut
  {NoStop}%
\bibitem [{\citenamefont {Bombelli}\ \emph {et~al.}(1986)\citenamefont
  {Bombelli}, \citenamefont {Koul}, \citenamefont {Lee},\ and\ \citenamefont
  {Sorkin}}]{SorkinArea1986}%
  \BibitemOpen
  \bibfield  {author} {\bibinfo {author} {\bibfnamefont {L.}~\bibnamefont
  {Bombelli}}, \bibinfo {author} {\bibfnamefont {R.~K.}\ \bibnamefont {Koul}},
  \bibinfo {author} {\bibfnamefont {J.}~\bibnamefont {Lee}}, \ and\ \bibinfo
  {author} {\bibfnamefont {R.~D.}\ \bibnamefont {Sorkin}},\ }\bibfield  {title}
  {\enquote {\bibinfo {title} {Quantum source of entropy for black holes},}\
  }\href {\doibase 10.1103/PhysRevD.34.373} {\bibfield  {journal} {\bibinfo
  {journal} {Phys. Rev. D}\ }\textbf {\bibinfo {volume} {34}},\ \bibinfo
  {pages} {373--383} (\bibinfo {year} {1986})}\BibitemShut {NoStop}%
\bibitem [{\citenamefont {Srednicki}(1993)}]{areaLaw1993}%
  \BibitemOpen
  \bibfield  {author} {\bibinfo {author} {\bibfnamefont {M.}~\bibnamefont
  {Srednicki}},\ }\bibfield  {title} {\enquote {\bibinfo {title} {Entropy and
  area},}\ }\href {\doibase 10.1103/PhysRevLett.71.666} {\bibfield  {journal}
  {\bibinfo  {journal} {Phys. Rev. Lett.}\ }\textbf {\bibinfo {volume} {71}},\
  \bibinfo {pages} {666--669} (\bibinfo {year} {1993})}\BibitemShut {NoStop}%
\bibitem [{\citenamefont {Eisert}\ \emph {et~al.}(2010)\citenamefont {Eisert},
  \citenamefont {Cramer},\ and\ \citenamefont {Plenio}}]{areaLawReview2010}%
  \BibitemOpen
  \bibfield  {author} {\bibinfo {author} {\bibfnamefont {J.}~\bibnamefont
  {Eisert}}, \bibinfo {author} {\bibfnamefont {M.}~\bibnamefont {Cramer}}, \
  and\ \bibinfo {author} {\bibfnamefont {M.~B.}\ \bibnamefont {Plenio}},\
  }\bibfield  {title} {\enquote {\bibinfo {title} {Colloquium: Area laws for
  the entanglement entropy},}\ }\href {\doibase 10.1103/RevModPhys.82.277}
  {\bibfield  {journal} {\bibinfo  {journal} {Rev. Mod. Phys.}\ }\textbf
  {\bibinfo {volume} {82}},\ \bibinfo {pages} {277--306} (\bibinfo {year}
  {2010})}\BibitemShut {NoStop}%
\bibitem [{\citenamefont {Witten}(2018)}]{witten}%
  \BibitemOpen
  \bibfield  {author} {\bibinfo {author} {\bibfnamefont {E.}~\bibnamefont
  {Witten}},\ }\bibfield  {title} {\enquote {\bibinfo {title} {{APS Medal for
  Exceptional Achievement in Research: Invited article on entanglement
  properties of quantum field theory}},}\ }\href {\doibase
  10.1103/RevModPhys.90.045003} {\bibfield  {journal} {\bibinfo  {journal}
  {Rev. Mod. Phys.}\ }\textbf {\bibinfo {volume} {90}},\ \bibinfo {pages}
  {045003} (\bibinfo {year} {2018})}\BibitemShut {NoStop}%
\bibitem [{\citenamefont {Hawking}(1974)}]{HawkingRadiation}%
  \BibitemOpen
  \bibfield  {author} {\bibinfo {author} {\bibfnamefont {S.~W.}\ \bibnamefont
  {Hawking}},\ }\bibfield  {title} {\enquote {\bibinfo {title} {{Black hole
  explosions}},}\ }\href {\doibase 10.1038/248030a0} {\bibfield  {journal}
  {\bibinfo  {journal} {Nature}\ }\textbf {\bibinfo {volume} {248}},\ \bibinfo
  {pages} {30--31} (\bibinfo {year} {1974})}\BibitemShut {NoStop}%
\bibitem [{\citenamefont {Hawking}(1975)}]{hawking_particle_1975}%
  \BibitemOpen
  \bibfield  {author} {\bibinfo {author} {\bibfnamefont {S.~W.}\ \bibnamefont
  {Hawking}},\ }\bibfield  {title} {\enquote {\bibinfo {title} {Particle
  creation by black holes},}\ }\href {\doibase 10.1007/BF02345020} {\bibfield
  {journal} {\bibinfo  {journal} {Communications in Mathematical Physics}\
  }\textbf {\bibinfo {volume} {43}},\ \bibinfo {pages} {199--220} (\bibinfo
  {year} {1975})}\BibitemShut {NoStop}%
\bibitem [{\citenamefont {Davies}(1975)}]{Davies1974}%
  \BibitemOpen
  \bibfield  {author} {\bibinfo {author} {\bibfnamefont {P.~C.~W.}\
  \bibnamefont {Davies}},\ }\bibfield  {title} {\enquote {\bibinfo {title}
  {{Scalar particle production in Schwarzschild and Rindler metrics}},}\ }\href
  {\doibase 10.1088/0305-4470/8/4/022} {\bibfield  {journal} {\bibinfo
  {journal} {J. Phys. A}\ }\textbf {\bibinfo {volume} {8}},\ \bibinfo {pages}
  {609--616} (\bibinfo {year} {1975})}\BibitemShut {NoStop}%
\bibitem [{\citenamefont {Page}(1993)}]{Page1993}%
  \BibitemOpen
  \bibfield  {author} {\bibinfo {author} {\bibfnamefont {D.~N.}\ \bibnamefont
  {Page}},\ }\bibfield  {title} {\enquote {\bibinfo {title} {Information in
  black hole radiation},}\ }\href {\doibase 10.1103/PhysRevLett.71.3743}
  {\bibfield  {journal} {\bibinfo  {journal} {Phys. Rev. Lett.}\ }\textbf
  {\bibinfo {volume} {71}},\ \bibinfo {pages} {3743--3746} (\bibinfo {year}
  {1993})}\BibitemShut {NoStop}%
\bibitem [{\citenamefont {Page}(2013)}]{Page2013}%
  \BibitemOpen
  \bibfield  {author} {\bibinfo {author} {\bibfnamefont {D.~N.}\ \bibnamefont
  {Page}},\ }\bibfield  {title} {\enquote {\bibinfo {title} {{Time dependence
  of Hawking radiation entropy}},}\ }\href {\doibase
  10.1088/1475-7516/2013/09/028} {\bibfield  {journal} {\bibinfo  {journal} {J.
  Cosmol. Astropart. Phys.}\ }\textbf {\bibinfo {volume} {2013}},\ \bibinfo
  {pages} {028--028} (\bibinfo {year} {2013})}\BibitemShut {NoStop}%
\bibitem [{\citenamefont {Penington}(2020)}]{Penington2020}%
  \BibitemOpen
  \bibfield  {author} {\bibinfo {author} {\bibfnamefont {G.}~\bibnamefont
  {Penington}},\ }\bibfield  {title} {\enquote {\bibinfo {title} {Entanglement
  wedge reconstruction and the information paradox},}\ }\href {\doibase
  10.1007/JHEP09(2020)002} {\bibfield  {journal} {\bibinfo  {journal} {J. High
  Energy Phys.}\ }\textbf {\bibinfo {volume} {2020}},\ \bibinfo {pages} {2}
  (\bibinfo {year} {2020})}\BibitemShut {NoStop}%
\bibitem [{\citenamefont {Reznik}\ \emph {et~al.}(2005)\citenamefont {Reznik},
  \citenamefont {Retzker},\ and\ \citenamefont {Silman}}]{reznik1}%
  \BibitemOpen
  \bibfield  {author} {\bibinfo {author} {\bibfnamefont {B.}~\bibnamefont
  {Reznik}}, \bibinfo {author} {\bibfnamefont {A.}~\bibnamefont {Retzker}}, \
  and\ \bibinfo {author} {\bibfnamefont {J.}~\bibnamefont {Silman}},\
  }\bibfield  {title} {\enquote {\bibinfo {title} {{Violating Bell's
  inequalities in vacuum}},}\ }\href
  {http://link.aps.org/abstract/PRA/v71/e042104} {\bibfield  {journal}
  {\bibinfo  {journal} {Phys. Rev. A}\ }\textbf {\bibinfo {volume} {71}},\
  \bibinfo {eid} {042104} (\bibinfo {year} {2005})}\BibitemShut {NoStop}%
\bibitem [{\citenamefont {Pozas-Kerstjens}\ and\ \citenamefont
  {Mart\'{i}n-Mart\'{i}nez}(2015)}]{Pozas-Kerstjens:2015}%
  \BibitemOpen
  \bibfield  {author} {\bibinfo {author} {\bibfnamefont {A.}~\bibnamefont
  {Pozas-Kerstjens}}\ and\ \bibinfo {author} {\bibfnamefont {E.}~\bibnamefont
  {Mart\'{i}n-Mart\'{i}nez}},\ }\bibfield  {title} {\enquote {\bibinfo {title}
  {Harvesting correlations from the quantum vacuum},}\ }\href {\doibase
  10.1103/PhysRevD.92.064042} {\bibfield  {journal} {\bibinfo  {journal} {Phys.
  Rev. D}\ }\textbf {\bibinfo {volume} {92}},\ \bibinfo {pages} {064042}
  (\bibinfo {year} {2015})}\BibitemShut {NoStop}%
\bibitem [{\citenamefont {Valentini}(1991)}]{Valentini1991}%
  \BibitemOpen
  \bibfield  {author} {\bibinfo {author} {\bibfnamefont {A.}~\bibnamefont
  {Valentini}},\ }\bibfield  {title} {\enquote {\bibinfo {title} {Non-local
  correlations in quantum electrodynamics},}\ }\href {\doibase
  http://dx.doi.org/10.1016/0375-9601(91)90952-5} {\bibfield  {journal}
  {\bibinfo  {journal} {Phys. Lett. A}\ }\textbf {\bibinfo {volume} {153}},\
  \bibinfo {pages} {321 -- 325} (\bibinfo {year} {1991})}\BibitemShut {NoStop}%
\bibitem [{\citenamefont {Reznik}(2003)}]{Reznik2003}%
  \BibitemOpen
  \bibfield  {author} {\bibinfo {author} {\bibfnamefont {B.}~\bibnamefont
  {Reznik}},\ }\bibfield  {title} {\enquote {\bibinfo {title} {Entanglement
  from the vacuum},}\ }\href@noop {} {\bibfield  {journal} {\bibinfo  {journal}
  {Foundations of Physics}\ }\textbf {\bibinfo {volume} {33}},\ \bibinfo
  {pages} {167--176} (\bibinfo {year} {2003})}\BibitemShut {NoStop}%
\bibitem [{\citenamefont {VerSteeg}\ and\ \citenamefont
  {Menicucci}(2009)}]{Nick}%
  \BibitemOpen
  \bibfield  {author} {\bibinfo {author} {\bibfnamefont {G.}~\bibnamefont
  {VerSteeg}}\ and\ \bibinfo {author} {\bibfnamefont {N.~C.}\ \bibnamefont
  {Menicucci}},\ }\bibfield  {title} {\enquote {\bibinfo {title} {Entangling
  power of an expanding universe},}\ }\href@noop {} {\bibfield  {journal}
  {\bibinfo  {journal} {Phys. Rev. D}\ }\textbf {\bibinfo {volume} {79}},\
  \bibinfo {pages} {044027} (\bibinfo {year} {2009})}\BibitemShut {NoStop}%
\bibitem [{\citenamefont {Mart\'{\i}n-Mart\'{\i}nez}\ \emph
  {et~al.}(2016)\citenamefont {Mart\'{\i}n-Mart\'{\i}nez}, \citenamefont
  {Smith},\ and\ \citenamefont {Terno}}]{topology}%
  \BibitemOpen
  \bibfield  {author} {\bibinfo {author} {\bibfnamefont {E.}~\bibnamefont
  {Mart\'{\i}n-Mart\'{\i}nez}}, \bibinfo {author} {\bibfnamefont {A.~R.~H.}\
  \bibnamefont {Smith}}, \ and\ \bibinfo {author} {\bibfnamefont {D.~R.}\
  \bibnamefont {Terno}},\ }\bibfield  {title} {\enquote {\bibinfo {title}
  {Spacetime structure and vacuum entanglement},}\ }\href {\doibase
  10.1103/PhysRevD.93.044001} {\bibfield  {journal} {\bibinfo  {journal} {Phys.
  Rev. D}\ }\textbf {\bibinfo {volume} {93}},\ \bibinfo {pages} {044001}
  (\bibinfo {year} {2016})}\BibitemShut {NoStop}%
\bibitem [{\citenamefont {Henderson}\ \emph {et~al.}(2018)\citenamefont
  {Henderson}, \citenamefont {Hennigar}, \citenamefont {Mann}, \citenamefont
  {Smith},\ and\ \citenamefont {Zhang}}]{Henderson_2018}%
  \BibitemOpen
  \bibfield  {author} {\bibinfo {author} {\bibfnamefont {L.~J.}\ \bibnamefont
  {Henderson}}, \bibinfo {author} {\bibfnamefont {R.~A.}\ \bibnamefont
  {Hennigar}}, \bibinfo {author} {\bibfnamefont {R.~B.}\ \bibnamefont {Mann}},
  \bibinfo {author} {\bibfnamefont {A.~R.~H.}\ \bibnamefont {Smith}}, \ and\
  \bibinfo {author} {\bibfnamefont {J.}~\bibnamefont {Zhang}},\ }\bibfield
  {title} {\enquote {\bibinfo {title} {Harvesting entanglement from the black
  hole vacuum},}\ }\href {\doibase 10.1088/1361-6382/aae27e} {\bibfield
  {journal} {\bibinfo  {journal} {Classical and Quantum Gravity}\ }\textbf
  {\bibinfo {volume} {35}},\ \bibinfo {pages} {21LT02} (\bibinfo {year}
  {2018})}\BibitemShut {NoStop}%
\bibitem [{\citenamefont {Gallock-Yoshimura}\ \emph {et~al.}(2021)\citenamefont
  {Gallock-Yoshimura}, \citenamefont {Tjoa},\ and\ \citenamefont
  {Mann}}]{KenMannSCHBH}%
  \BibitemOpen
  \bibfield  {author} {\bibinfo {author} {\bibfnamefont {K.}~\bibnamefont
  {Gallock-Yoshimura}}, \bibinfo {author} {\bibfnamefont {E.}~\bibnamefont
  {Tjoa}}, \ and\ \bibinfo {author} {\bibfnamefont {R.~B.}\ \bibnamefont
  {Mann}},\ }\bibfield  {title} {\enquote {\bibinfo {title} {Harvesting
  entanglement with detectors freely falling into a black hole},}\ }\href
  {\doibase 10.1103/PhysRevD.104.025001} {\bibfield  {journal} {\bibinfo
  {journal} {Phys. Rev. D}\ }\textbf {\bibinfo {volume} {104}},\ \bibinfo
  {pages} {025001} (\bibinfo {year} {2021})}\BibitemShut {NoStop}%
\bibitem [{\citenamefont {Cliche}\ and\ \citenamefont
  {Kempf}(2011)}]{cliche2011vacuum}%
  \BibitemOpen
  \bibfield  {author} {\bibinfo {author} {\bibfnamefont {M.}~\bibnamefont
  {Cliche}}\ and\ \bibinfo {author} {\bibfnamefont {A.}~\bibnamefont {Kempf}},\
  }\bibfield  {title} {\enquote {\bibinfo {title} {Vacuum entanglement
  enhancement by a weak gravitational field},}\ }\href@noop {} {\bibfield
  {journal} {\bibinfo  {journal} {Physical Review D}\ }\textbf {\bibinfo
  {volume} {83}},\ \bibinfo {pages} {045019} (\bibinfo {year}
  {2011})}\BibitemShut {NoStop}%
\bibitem [{\citenamefont {Jonsson}\ \emph {et~al.}(2020)\citenamefont
  {Jonsson}, \citenamefont {Aruquipa}, \citenamefont {Casals}, \citenamefont
  {Kempf},\ and\ \citenamefont
  {{Mart{\'i}n-Mart{\'i}nez}}}]{jonssonCommunicationQuantumFields2020}%
  \BibitemOpen
  \bibfield  {author} {\bibinfo {author} {\bibfnamefont {R.~H.}\ \bibnamefont
  {Jonsson}}, \bibinfo {author} {\bibfnamefont {D.~Q.}\ \bibnamefont
  {Aruquipa}}, \bibinfo {author} {\bibfnamefont {M.}~\bibnamefont {Casals}},
  \bibinfo {author} {\bibfnamefont {A.}~\bibnamefont {Kempf}}, \ and\ \bibinfo
  {author} {\bibfnamefont {E.}~\bibnamefont {{Mart{\'i}n-Mart{\'i}nez}}},\
  }\bibfield  {title} {\enquote {\bibinfo {title} {Communication through
  quantum fields near a black hole},}\ }\href {\doibase
  10.1103/PhysRevD.101.125005} {\bibfield  {journal} {\bibinfo  {journal}
  {Physical Review D}\ }\textbf {\bibinfo {volume} {101}},\ \bibinfo {pages}
  {125005} (\bibinfo {year} {2020})}\BibitemShut {NoStop}%
\bibitem [{\citenamefont
  {Candelas}(1980)}]{candelasVacuumPolarizationSchwarzschild1980}%
  \BibitemOpen
  \bibfield  {author} {\bibinfo {author} {\bibfnamefont {P.}~\bibnamefont
  {Candelas}},\ }\bibfield  {title} {\enquote {\bibinfo {title} {Vacuum
  polarization in {{Schwarzschild}} spacetime},}\ }\href {\doibase
  10.1103/PhysRevD.21.2185} {\bibfield  {journal} {\bibinfo  {journal}
  {Physical Review D}\ }\textbf {\bibinfo {volume} {21}},\ \bibinfo {pages}
  {2185--2202} (\bibinfo {year} {1980})}\BibitemShut {NoStop}%
\bibitem [{\citenamefont {Boulware}(1975)}]{boulware1975spin}%
  \BibitemOpen
  \bibfield  {author} {\bibinfo {author} {\bibfnamefont {D.~G.}\ \bibnamefont
  {Boulware}},\ }\bibfield  {title} {\enquote {\bibinfo {title} {Spin-1 2
  quantum field theory in {S}chwarzschild space},}\ }\href@noop {} {\bibfield
  {journal} {\bibinfo  {journal} {Phys. Rev. D}\ }\textbf {\bibinfo {volume}
  {12}},\ \bibinfo {pages} {350} (\bibinfo {year} {1975})}\BibitemShut
  {NoStop}%
\bibitem [{\citenamefont {Unruh}(1976)}]{Unruh1976}%
  \BibitemOpen
  \bibfield  {author} {\bibinfo {author} {\bibfnamefont {W.~G.}\ \bibnamefont
  {Unruh}},\ }\bibfield  {title} {\enquote {\bibinfo {title} {Notes on
  black-hole evaporation},}\ }\href {\doibase 10.1103/PhysRevD.14.870}
  {\bibfield  {journal} {\bibinfo  {journal} {Phys. Rev. D}\ }\textbf {\bibinfo
  {volume} {14}},\ \bibinfo {pages} {870--892} (\bibinfo {year}
  {1976})}\BibitemShut {NoStop}%
\bibitem [{\citenamefont {Hartle}\ and\ \citenamefont
  {Hawking}(1976)}]{hartle1976path}%
  \BibitemOpen
  \bibfield  {author} {\bibinfo {author} {\bibfnamefont {J.~B.}\ \bibnamefont
  {Hartle}}\ and\ \bibinfo {author} {\bibfnamefont {S.~W.}\ \bibnamefont
  {Hawking}},\ }\bibfield  {title} {\enquote {\bibinfo {title} {Path-integral
  derivation of black-hole radiance},}\ }\href@noop {} {\bibfield  {journal}
  {\bibinfo  {journal} {Phys. Rev. D}\ }\textbf {\bibinfo {volume} {13}},\
  \bibinfo {pages} {2188} (\bibinfo {year} {1976})}\BibitemShut {NoStop}%
\bibitem [{\citenamefont {Hollands}\ and\ \citenamefont
  {Wald}(2015)}]{HOLLANDS20151}%
  \BibitemOpen
  \bibfield  {author} {\bibinfo {author} {\bibfnamefont {S.}~\bibnamefont
  {Hollands}}\ and\ \bibinfo {author} {\bibfnamefont {R.~M.}\ \bibnamefont
  {Wald}},\ }\bibfield  {title} {\enquote {\bibinfo {title} {Quantum fields in
  curved spacetime},}\ }\href {\doibase
  https://doi.org/10.1016/j.physrep.2015.02.001} {\bibfield  {journal}
  {\bibinfo  {journal} {Physics Reports}\ }\textbf {\bibinfo {volume} {574}},\
  \bibinfo {pages} {1--35} (\bibinfo {year} {2015})},\ \bibinfo {note} {quantum
  fields in curved spacetime}\BibitemShut {NoStop}%
\bibitem [{\citenamefont {DeWitt}\ and\ \citenamefont
  {Brehme}(1960)}]{DeWitt:1960}%
  \BibitemOpen
  \bibfield  {author} {\bibinfo {author} {\bibfnamefont {B.~S.}\ \bibnamefont
  {DeWitt}}\ and\ \bibinfo {author} {\bibfnamefont {R.~W.}\ \bibnamefont
  {Brehme}},\ }\bibfield  {title} {\enquote {\bibinfo {title} {{Radiation
  damping in a gravitational field}},}\ }\href {\doibase
  10.1016/0003-4916(60)90030-0} {\bibfield  {journal} {\bibinfo  {journal}
  {Ann. Phys.}\ }\textbf {\bibinfo {volume} {9}},\ \bibinfo {pages} {220--259}
  (\bibinfo {year} {1960})}\BibitemShut {NoStop}%
\bibitem [{\citenamefont {Hadamard}(1923)}]{Hadamard}%
  \BibitemOpen
  \bibfield  {author} {\bibinfo {author} {\bibfnamefont {J.}~\bibnamefont
  {Hadamard}},\ }\href@noop {} {\emph {\bibinfo {title} {Lectures on Cauchy's
  Problem in Linear Partial Differential Equations}}}\ (\bibinfo  {publisher}
  {Dover Publications},\ \bibinfo {year} {1923})\BibitemShut {NoStop}%
\bibitem [{\citenamefont {Garabedian}(1998)}]{Garabedian}%
  \BibitemOpen
  \bibfield  {author} {\bibinfo {author} {\bibfnamefont {P.~R.}\ \bibnamefont
  {Garabedian}},\ }\href@noop {} {\emph {\bibinfo {title} {Partial Differential
  Equations}}}\ (\bibinfo  {publisher} {Chelsea Pub Co},\ \bibinfo {address}
  {New York},\ \bibinfo {year} {1998})\BibitemShut {NoStop}%
\bibitem [{\citenamefont {Ikawa}(2000)}]{Ikawa}%
  \BibitemOpen
  \bibfield  {author} {\bibinfo {author} {\bibfnamefont {M.}~\bibnamefont
  {Ikawa}},\ }\href@noop {} {\emph {\bibinfo {title} {Hyperbolic partial
  differential equations and wave phenomena. Iwanami series in modern
  mathematics. Translations of mathematical monographs}}}\ (\bibinfo
  {publisher} {American Mathematical Soc.},\ \bibinfo {address} {Providence},\
  \bibinfo {year} {2000})\BibitemShut {NoStop}%
\bibitem [{\citenamefont {Buss}\ and\ \citenamefont
  {Casals}(2018)}]{BUSS2018168}%
  \BibitemOpen
  \bibfield  {author} {\bibinfo {author} {\bibfnamefont {C.}~\bibnamefont
  {Buss}}\ and\ \bibinfo {author} {\bibfnamefont {M.}~\bibnamefont {Casals}},\
  }\bibfield  {title} {\enquote {\bibinfo {title} {Quantum correlator outside a
  {{Schwarzschild}} black hole},}\ }\href {\doibase
  https://doi.org/10.1016/j.physletb.2017.11.048} {\bibfield  {journal}
  {\bibinfo  {journal} {Phys. Lett. B}\ }\textbf {\bibinfo {volume} {776}},\
  \bibinfo {pages} {168 -- 173} (\bibinfo {year} {2018})}\BibitemShut {NoStop}%
\bibitem [{\citenamefont {Casals}\ and\ \citenamefont
  {Nolan}(2016)}]{casals2016global}%
  \BibitemOpen
  \bibfield  {author} {\bibinfo {author} {\bibfnamefont {M.}~\bibnamefont
  {Casals}}\ and\ \bibinfo {author} {\bibfnamefont {B.}~\bibnamefont {Nolan}},\
  }\href@noop {} {\enquote {\bibinfo {title} {Global {{Hadamard}} form for the
  green function in {{Schwarzschild}} space-time},}\ } (\bibinfo {year}
  {2016}),\ \Eprint {http://arxiv.org/abs/arXiv:1606.03075} {arXiv:1606.03075}
  \BibitemShut {NoStop}%
\bibitem [{\citenamefont {DeWitt}(1979)}]{dewittQuantumGravityNew1979}%
  \BibitemOpen
  \bibfield  {author} {\bibinfo {author} {\bibfnamefont {B.~S.}\ \bibnamefont
  {DeWitt}},\ }\bibfield  {title} {\enquote {\bibinfo {title} {Quantum gravity:
  The new synthesis},}\ }in\ \href@noop {} {\emph {\bibinfo {booktitle}
  {General Relativity : An {{Einstein}} Centenary Survey}}},\ \bibinfo {editor}
  {edited by\ \bibinfo {editor} {\bibfnamefont {S.}~\bibnamefont {Hawking}}\
  and\ \bibinfo {editor} {\bibfnamefont {W.}~\bibnamefont {Israel}}}\ (\bibinfo
   {publisher} {{Cambridge University Press}},\ \bibinfo {address} {{Cambridge
  Eng; New York}},\ \bibinfo {year} {1979})\ p.\ \bibinfo {pages}
  {680}\BibitemShut {NoStop}%
\bibitem [{\citenamefont {Mart\'{\i}n-Mart\'{\i}nez}\ \emph
  {et~al.}(2021)\citenamefont {Mart\'{\i}n-Mart\'{\i}nez}, \citenamefont
  {Perche},\ and\ \citenamefont {Torres}}]{Martin-MartinezPercheSouza2}%
  \BibitemOpen
  \bibfield  {author} {\bibinfo {author} {\bibfnamefont {E.}~\bibnamefont
  {Mart\'{\i}n-Mart\'{\i}nez}}, \bibinfo {author} {\bibfnamefont {T.~R.}\
  \bibnamefont {Perche}}, \ and\ \bibinfo {author} {\bibfnamefont {B.~d.
  S.~L.}\ \bibnamefont {Torres}},\ }\bibfield  {title} {\enquote {\bibinfo
  {title} {Broken covariance of particle detector models in relativistic
  quantum information},}\ }\href {\doibase 10.1103/PhysRevD.103.025007}
  {\bibfield  {journal} {\bibinfo  {journal} {Phys. Rev. D}\ }\textbf {\bibinfo
  {volume} {103}},\ \bibinfo {pages} {025007} (\bibinfo {year}
  {2021})}\BibitemShut {NoStop}%
\bibitem [{\citenamefont {de~Ram\'on}\ \emph {et~al.}(2021)\citenamefont
  {de~Ram\'on}, \citenamefont {Papageorgiou},\ and\ \citenamefont
  {Mart\'{\i}n-Mart\'{\i}nez}}]{deRamonPapageorgiumartin-martinez}%
  \BibitemOpen
  \bibfield  {author} {\bibinfo {author} {\bibfnamefont {J.}~\bibnamefont
  {de~Ram\'on}}, \bibinfo {author} {\bibfnamefont {M.}~\bibnamefont
  {Papageorgiou}}, \ and\ \bibinfo {author} {\bibfnamefont {E.}~\bibnamefont
  {Mart\'{\i}n-Mart\'{\i}nez}},\ }\bibfield  {title} {\enquote {\bibinfo
  {title} {Relativistic causality in particle detector models:
  Faster-than-light signaling and impossible measurements},}\ }\href {\doibase
  10.1103/PhysRevD.103.085002} {\bibfield  {journal} {\bibinfo  {journal}
  {Phys. Rev. D}\ }\textbf {\bibinfo {volume} {103}},\ \bibinfo {pages}
  {085002} (\bibinfo {year} {2021})}\BibitemShut {NoStop}%
\bibitem [{\citenamefont
  {{Mart{\'i}n-Mart{\'i}nez}}(2015)}]{martin-martinezCausalityIssuesParticle2015}%
  \BibitemOpen
  \bibfield  {author} {\bibinfo {author} {\bibfnamefont {E.}~\bibnamefont
  {{Mart{\'i}n-Mart{\'i}nez}}},\ }\bibfield  {title} {\enquote {\bibinfo
  {title} {Causality issues of particle detector models in {{QFT}} and quantum
  optics},}\ }\href {\doibase 10.1103/PhysRevD.92.104019} {\bibfield  {journal}
  {\bibinfo  {journal} {Physical Review D}\ }\textbf {\bibinfo {volume} {92}},\
  \bibinfo {pages} {104019} (\bibinfo {year} {2015})}\BibitemShut {NoStop}%
\bibitem [{\citenamefont {Mart\'{\i}n-Mart\'{\i}nez}\ \emph
  {et~al.}(2020)\citenamefont {Mart\'{\i}n-Mart\'{\i}nez}, \citenamefont
  {Perche},\ and\ \citenamefont
  {de~S.~L.~Torres}}]{Martin-MartinezPercheSouza1}%
  \BibitemOpen
  \bibfield  {author} {\bibinfo {author} {\bibfnamefont {E.}~\bibnamefont
  {Mart\'{\i}n-Mart\'{\i}nez}}, \bibinfo {author} {\bibfnamefont {T.~R.}\
  \bibnamefont {Perche}}, \ and\ \bibinfo {author} {\bibfnamefont
  {B.}~\bibnamefont {de~S.~L.~Torres}},\ }\bibfield  {title} {\enquote
  {\bibinfo {title} {General relativistic quantum optics: Finite-size particle
  detector models in curved spacetimes},}\ }\href {\doibase
  10.1103/PhysRevD.101.045017} {\bibfield  {journal} {\bibinfo  {journal}
  {Phys. Rev. D}\ }\textbf {\bibinfo {volume} {101}},\ \bibinfo {pages}
  {045017} (\bibinfo {year} {2020})}\BibitemShut {NoStop}%
\bibitem [{\citenamefont {Lopp}\ and\ \citenamefont
  {Mart\'{i}n-Mart\'{i}nez}(2021)}]{Richard}%
  \BibitemOpen
  \bibfield  {author} {\bibinfo {author} {\bibfnamefont {R.}~\bibnamefont
  {Lopp}}\ and\ \bibinfo {author} {\bibfnamefont {E.}~\bibnamefont
  {Mart\'{i}n-Mart\'{i}nez}},\ }\bibfield  {title} {\enquote {\bibinfo {title}
  {Quantum delocalization, gauge, and quantum optics: Light-matter interaction
  in relativistic quantum information},}\ }\href {\doibase
  10.1103/PhysRevA.103.013703} {\bibfield  {journal} {\bibinfo  {journal}
  {Phys. Rev. A}\ }\textbf {\bibinfo {volume} {103}},\ \bibinfo {pages}
  {013703} (\bibinfo {year} {2021})}\BibitemShut {NoStop}%
\bibitem [{\citenamefont {{Pozas-Kerstjens}}\ and\ \citenamefont
  {{Martin-Martinez}}(2015)}]{pozas-kerstjensHarvestingCorrelationsQuantum2015}%
  \BibitemOpen
  \bibfield  {author} {\bibinfo {author} {\bibfnamefont {A.}~\bibnamefont
  {{Pozas-Kerstjens}}}\ and\ \bibinfo {author} {\bibfnamefont {E.}~\bibnamefont
  {{Martin-Martinez}}},\ }\bibfield  {title} {\enquote {\bibinfo {title}
  {Harvesting correlations from the quantum vacuum},}\ }\href {\doibase
  10.1103/PhysRevD.92.064042} {\bibfield  {journal} {\bibinfo  {journal}
  {Physical Review D}\ }\textbf {\bibinfo {volume} {92}},\ \bibinfo {pages}
  {064042} (\bibinfo {year} {2015})}\BibitemShut {NoStop}%
\bibitem [{\citenamefont {{Martin-Martinez}}\ and\ \citenamefont
  {{Rodriguez-Lopez}}(2018)}]{martin-martinezRelativisticQuantumOptics2018}%
  \BibitemOpen
  \bibfield  {author} {\bibinfo {author} {\bibfnamefont {E.}~\bibnamefont
  {{Martin-Martinez}}}\ and\ \bibinfo {author} {\bibfnamefont {P.}~\bibnamefont
  {{Rodriguez-Lopez}}},\ }\bibfield  {title} {\enquote {\bibinfo {title}
  {Relativistic {{Quantum Optics}}: {{On}} the relativistic invariance of the
  light-matter interaction models},}\ }\href {\doibase
  10.1103/PhysRevD.97.105026} {\bibfield  {journal} {\bibinfo  {journal}
  {Physical Review D}\ }\textbf {\bibinfo {volume} {97}},\ \bibinfo {pages}
  {105026} (\bibinfo {year} {2018})}\BibitemShut {NoStop}%
\bibitem [{\citenamefont {Tjoa}\ and\ \citenamefont
  {{Mart{\'i}n-Mart{\'i}nez}}(2021)}]{tjoaWhenEntanglementHarvesting2021}%
  \BibitemOpen
  \bibfield  {author} {\bibinfo {author} {\bibfnamefont {E.}~\bibnamefont
  {Tjoa}}\ and\ \bibinfo {author} {\bibfnamefont {E.}~\bibnamefont
  {{Mart{\'i}n-Mart{\'i}nez}}},\ }\bibfield  {title} {\enquote {\bibinfo
  {title} {When entanglement harvesting is not really harvesting},}\ }\href
  {\doibase 10.1103/PhysRevD.104.125005} {\bibfield  {journal} {\bibinfo
  {journal} {Physical Review D}\ }\textbf {\bibinfo {volume} {104}},\ \bibinfo
  {pages} {125005} (\bibinfo {year} {2021})}\BibitemShut {NoStop}%
\bibitem [{\citenamefont {{Wolfram Research, Inc.}}()}]{Mathematica}%
  \BibitemOpen
  \bibfield  {author} {\bibinfo {author} {\bibnamefont {{Wolfram Research,
  Inc.}}},\ }\href {https://www.wolfram.com/mathematica} {\enquote {\bibinfo
  {title} {Mathematica, {V}ersion 13.2},}\ }\bibinfo {note} {Champaign, IL,
  2022}\BibitemShut {NoStop}%
\bibitem [{BHP()}]{BHPToolkit}%
  \BibitemOpen
  \href@noop {} {\enquote {\bibinfo {title} {{Black Hole Perturbation
  Toolkit}},}\ }\bibinfo {howpublished}
  {(\href{http://bhptoolkit.org/}{bhptoolkit.org})}\BibitemShut {NoStop}%
\bibitem [{\citenamefont {Leaver}(1986)}]{Leaver1986}%
  \BibitemOpen
  \bibfield  {author} {\bibinfo {author} {\bibfnamefont {E.~W.}\ \bibnamefont
  {Leaver}},\ }\bibfield  {title} {\enquote {\bibinfo {title} {{Solutions to a
  generalized spheroidal wave equation: Teukolsky's equations in general
  relativity, and the two‐center problem in molecular quantum mechanics}},}\
  }\href {\doibase 10.1063/1.527130} {\bibfield  {journal} {\bibinfo  {journal}
  {Journal of Mathematical Physics}\ }\textbf {\bibinfo {volume} {27}},\
  \bibinfo {pages} {1238--1265} (\bibinfo {year} {1986})}\BibitemShut {NoStop}%
\bibitem [{\citenamefont {Sasaki}\ and\ \citenamefont
  {Tagoshi}(2003)}]{Sasaki2003}%
  \BibitemOpen
  \bibfield  {author} {\bibinfo {author} {\bibfnamefont {M.}~\bibnamefont
  {Sasaki}}\ and\ \bibinfo {author} {\bibfnamefont {H.}~\bibnamefont
  {Tagoshi}},\ }\bibfield  {title} {\enquote {\bibinfo {title} {{Analytic Black
  Hole Perturbation Approach to Gravitational Radiation}},}\ }\href {\doibase
  10.12942/lrr-2003-6} {\bibfield  {journal} {\bibinfo  {journal} {Living
  Reviews in Relativity}\ }\textbf {\bibinfo {volume} {6}},\ \bibinfo {pages}
  {6} (\bibinfo {year} {2003})}\BibitemShut {NoStop}%
\bibitem [{\citenamefont {Levi}\ and\ \citenamefont {Ori}(2016)}]{Levi2016}%
  \BibitemOpen
  \bibfield  {author} {\bibinfo {author} {\bibfnamefont {A.}~\bibnamefont
  {Levi}}\ and\ \bibinfo {author} {\bibfnamefont {A.}~\bibnamefont {Ori}},\
  }\bibfield  {title} {\enquote {\bibinfo {title} {Mode-sum regularization of
  $⟨{\ensuremath{\phi}}^{2}⟩$ in the angular-splitting method},}\ }\href
  {\doibase 10.1103/PhysRevD.94.044054} {\bibfield  {journal} {\bibinfo
  {journal} {Phys. Rev. D}\ }\textbf {\bibinfo {volume} {94}},\ \bibinfo
  {pages} {044054} (\bibinfo {year} {2016})}\BibitemShut {NoStop}%
\end{thebibliography}%
\end{document}